\documentclass[reprint,prb,floatfix,superscriptaddress]{revtex4-1} 

\usepackage[utf8]{inputenc}   
\usepackage[T1]{fontenc}      
\usepackage[english]{babel}                             
\usepackage{amsmath,amssymb}
\usepackage{float}
\usepackage[]{graphicx} 
\usepackage{array}
\usepackage{subfigure}
\usepackage[pdftex]{hyperref}
\usepackage{miller} 
\usepackage{multirow}
\usepackage{mathrsfs}
\usepackage{dcolumn} 

\newcommand{\etal}{\textit{et al. }}
\newcommand{\ai}{\textit{ab initio }}
\newcommand{\Ai}{\textit{Ab initio }}

\newcommand{\mrm}[1]{\mathrm{#1}}

\newcommand{\burger}[2]{$\vec{b}_{#1}^\mathrm{#2}$}
\newcommand{\burg}[1]{$\vec{b}_{#1}$}

\begin{document}

\title{Atomic scale modeling of twinning disconnections in zirconium}
\author{Olivier MacKain}
\affiliation{DEN-Service de Recherches de Métallurgie Physique, CEA, Université Paris-Saclay, F-91191, Gif-sur-Yvette, France}
\author{Maeva Cottura}
\affiliation{DEN-Service de Recherches de Métallurgie Physique, CEA, Université Paris-Saclay, F-91191, Gif-sur-Yvette, France}
\author{David Rodney}
\affiliation{Institut Lumière Matière, CNRS-Université Claude Bernard Lyon 1, F-69622 Villeurbanne, France}
\author{Emmanuel Clouet}
\email[Corresponding author: ]{emmanuel.clouet@cea.fr}
\affiliation{DEN-Service de Recherches de Métallurgie Physique, CEA, Université Paris-Saclay, F-91191, Gif-sur-Yvette, France}

\begin{abstract}
	Twin growth in hexagonal close-packed zirconium is investigated at the atomic scale
	by modeling the various disconnections that can exist on twin boundaries.
	Thanks to a coupling with elasticity theory, 
	core energies are extracted from atomistic simulations 
	and the formation energy of isolated disconnection dipoles is defined. 
	For twin systems where several disconnections can exist, 
	because of this core contribution,
	the most stable disconnection is not always the one with the smallest Burgers vector.
	Crystallographic parameters of the disconnection with the lowest formation energy 
	correlate well with twin modes observed experimentally.
	On the other hand, disconnection migration, characterized here by computing their migration energy and Peierls stress, does not appear critical 
	for twin mode selection.
\end{abstract}

\maketitle

\section {Introduction}
\label{sec:Introduction}

Twinning is a necessary deformation mode in crystals where only a limited number 
of dislocation slip systems can be activated. 
This is the case in hexagonal close-packed (hcp) metals \cite{Partridge1967,Christian1995}
where dislocations with $1/3\,\hkl<1-210>$ Burgers vectors
are the main carriers of plastic deformation\cite{Hirth1982}
but cannot accommodate any strain along the \hkl<c> axis.
This can be done only through the activation of twinning 
or the glide of dislocations with $1/3\,\hkl<1-213>$ Burgers vectors.
Whereas $1/3\,\hkl<1-213>$ dislocations account for plastic deformation at fairly high temperatures, 
twinning becomes the dominant mechanism at low temperature or high strain rate.\cite{Yoo1981}
While the crystallography of twinning is well asserted,\cite{Christian1995} 
the mechanisms controlling twin formation and growth
are still the object of active research.\cite{Beyerlein2014}
Twinning can be divided in three steps: 
twin nucleation, lateral propagation and thickening.

This article focuses on the last step, 
where the motion of the twin boundary leads to thickening.
This motion occurs thanks to disconnections,\cite{Christian1995,Hirth2016}
i.e. interface dislocations with a step character 
gliding along the twin boundary to propagate locally
the twin in the parent crystal.
The presence of disconnections on the twin boundary can result 
from the interaction of the boundary with dislocations coming from the bulk crystal
\cite{Serra1995,Serra1996,Wang2012,Kadiri2015,Fan2015,Fan2016,Wang2016}
or from nucleation under the action of the applied stress and temperature.
\cite{Ghazisaeidi2014,Luque2014}
Whether twin thickening is controlled by the formation or the migration of the disconnections 
is not clearly known.  
In hcp metals, although several disconnections with different Burgers vectors and / or 
different step heights are usually possible on a twin boundary,\cite{Serra1988,Serra1991,Bacon2002} 
only a single twinning mode, corresponding to a given shear direction and intensity, 
is reported for each twin system.\cite{Partridge1967,Christian1995}
One can then wonder if the selection of this mode is driven by 
a competition between the formation of the different possible disconnections
or between their migration.
Answering such a question requires a description at the atomic scale.
Although crystallography and elasticity theory can rationalize 
many properties of disconnections, the core region of the line defects
controls their migration and has a non negligible contribution to the defect formation energy,
sometimes counterbalancing the elastic contribution as shown below.

In this perspective, disconnections are modeled here at the atomic scale in Zr, 
an hcp transition metal of upmost technological interest in the nuclear industry.\cite{Lemaignan2012}
The purpose is to  extract from these atomistic simulations the key quantities
describing disconnection formation and migration, 
and to confront these results with the twinning modes reported experimentally.
The developed method proposes to extract the disconnection core energy from atomistic simulations using small simulation cells and to calculate the elastic energy thanks to an analogy between a disconnection dipole and an Eshelby inclusion,
taking full account of the elastic anisotropy and inhomogeneity.
Small computation cells are used in order to later perform \ai calculations but for the moment, this approach is validated using a many-body potential of the embedded-atom method (EAM) type.
The adequacy of the EAM potential is first assessed by comparison with \ai calculations of perfect twin boundaries.
All possible disconnections that have been proposed for the four different twin systems 
reported in Zr\cite{Yoo1991} are then modeled.
The formation energies of disconnection dipoles are then defined considering both core and elastic contributions. 
The disconnection migration is finally studied by calculating their migration energy and Peierls stress.

\section{Perfect twin boundaries}
\label{sec:perfect_twins}
The empirical interatomic potential is validated by comparing the structures and energies of perfect twin boundaries in hcp Zr with \ai calculations.
The four twin systems that can be activated in Zr,\cite{Yoo1991} corresponding to \hkl{10-11}, \hkl{11-22}, \hkl{10-12} and \hkl{11-21} twin boundaries, are considered.

\subsection{Methods}
\label{subsec:Methods}
The atomic interactions are described using the EAM potential developed by Mendelev and Ackland for zirconium and referred to as \#3 in reference \onlinecite{Mendelev2007}.
This potential, developed for bulk hcp Zr with a special emphasis on stacking faults controlling dislocation dissociation, has already been shown to give a reliable description of Zr plasticity.\cite{Khater2010,Clouet2012,Chaari2014,Chaari2014a,Lu2015a,Lu2015,Szewc2016}

\Ai calculations were performed with the VASP code, \cite{Kresse1996}
using the Perdew-Becke-Erzenhof \cite{Perdew1996} parameterization for the GGA exchange and correlation functional.
The interactions between the core and outer electrons are modeled through the projector augmented wave approximation, considering valence (5s$^2$ 4d$^2$) and semicore electrons (4s$^2$ 4p$^6$) in the outer shell.
The plane wave cutoff energy is set at 460 eV.
The Brillouin zone is sampled with a regular $\Gamma$-centered mesh corresponding to $20\times20\times9$ $k$-points for the primitive hcp crystal cell.
The electronic density of states is integrated using the Methfessel-Paxton broadening function with a smearing parameter of 0.1 eV.
Atomic positions are relaxed until the force on each atom is less than $5\ 10^{-3}$\,eV\,\AA$^{-1}$.

Twin boundaries are modeled with periodic boundary conditions in all directions, thus introducing two twin planes in the simulation cell.
A minimum number of atomic planes has to be kept between the two twin planes to limit their interactions.
Relaxation perpendicular to the twin boundary is considered by allowing an increase or a decrease of the periodicity vector in this direction, with the corresponding displacement $\delta_{z}$ localized in the planes adjacent to  the twin boundaries.\cite{Kumar2015}
Such a relaxation improves the convergence of the twin energy with the number of atomic planes in the simulation cell.
For \ai calculations, cells containing nine atomic planes between twin planes are used leading to a convergence better than  5 mJ\,m$^{-2}$ on twin energies.

\subsection{Atomic structure}

\begin{figure}[!b]  
		\subfigure[\hkl{10-11}]{
		\includegraphics[width=.46\linewidth]{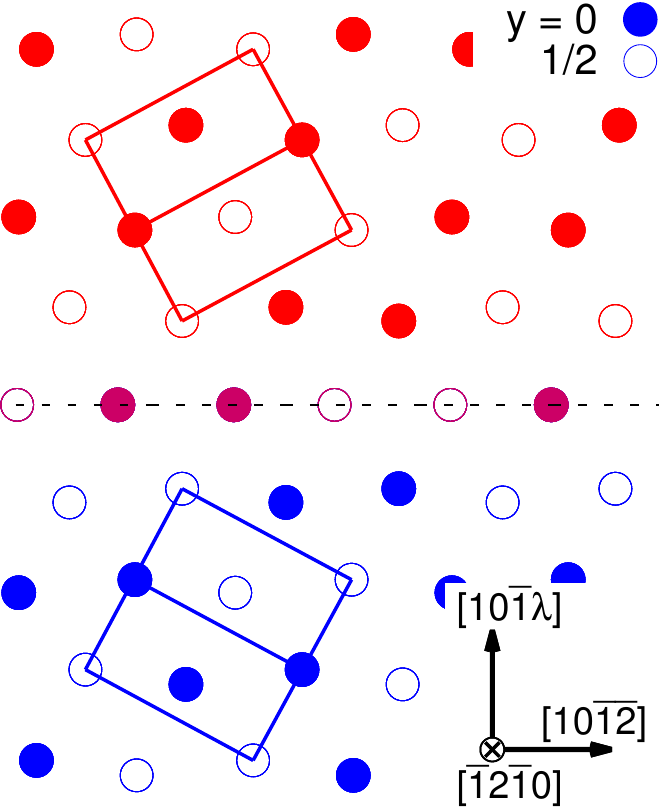}
		}
		\hfill
		\subfigure[\hkl{11-22}]{
		\includegraphics[width=.46\linewidth]{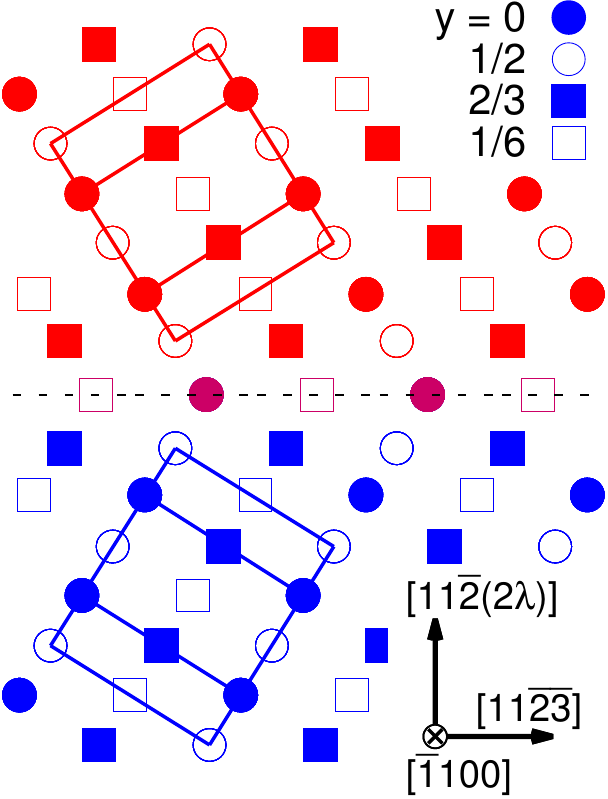}
		}
		\subfigure[\hkl{10-12}]{
		\includegraphics[width=.46\linewidth]{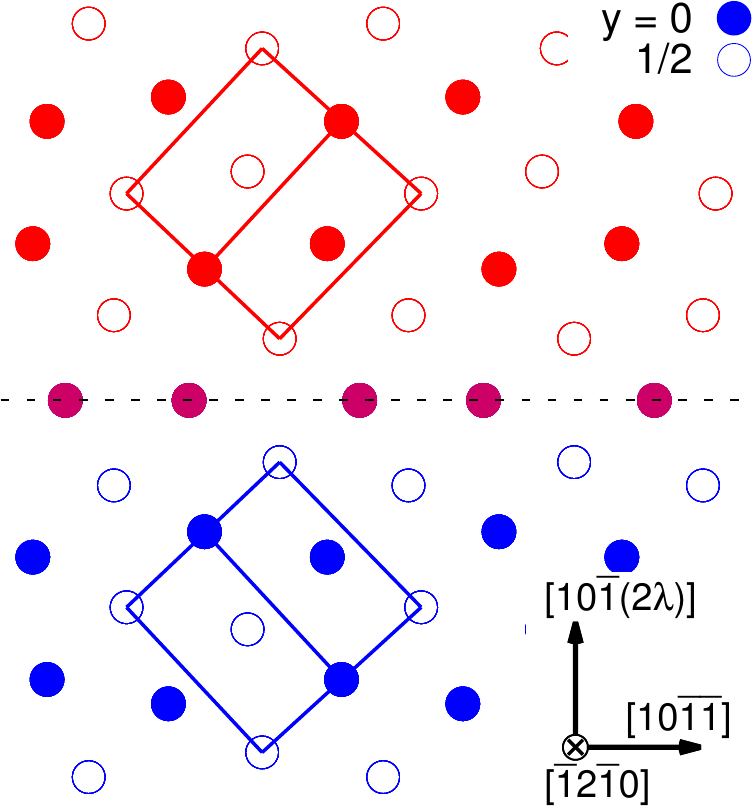}
		}
		\hfill
		\subfigure[\hkl{11-21}]{
		\includegraphics[width=.46\linewidth]{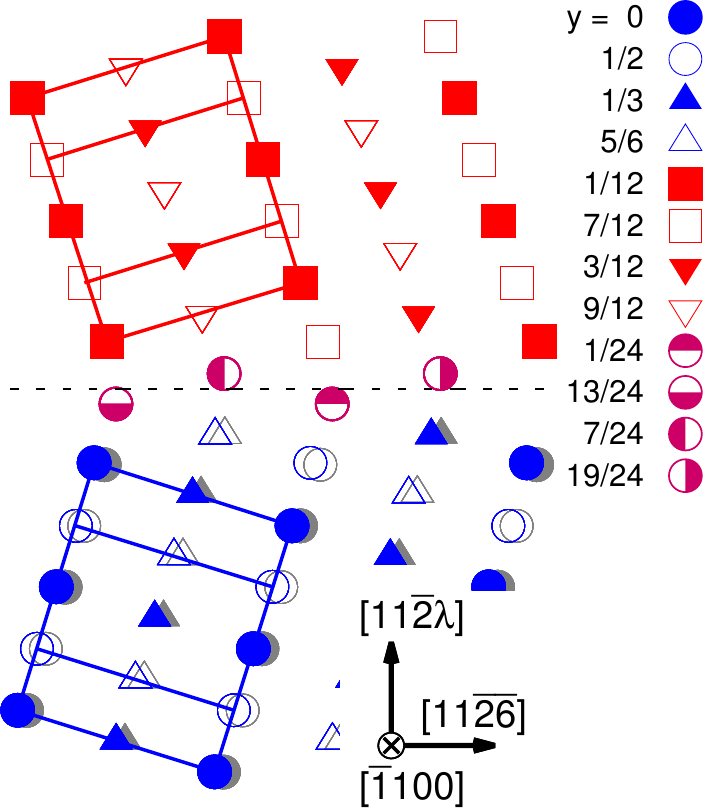}
		}
	\caption{Structures of the relaxed twin boundaries predicted by the EAM potential.
	The different symbols refer to the depth $y$ of the atoms along the periodicity vector normal to the figure plane.
	The color refers to atoms of the parent (blue), the twin crystal (red) or the twin plane (purple).
	(d) For \hkl{11-21} twin system, the grey symbols show the positions of perfectly symmetric twin boundary.
	The direction perpendicular to the twin boundary $z$ is defined through a factor $\lambda=3/(2\gamma^2)$, where $\gamma$ is the $c/a$ ratio of the hcp crystal.\cite{Frank1965}}
	\label{fig:Structures_EAM}
\end{figure}

Twinning operations can either correspond to a reflection in the twinning plane $K_1$ (Type I), or a rotation of $\pi$ along the twinning direction $\eta_1$ (Type II).\cite{Christian1995}
\hkl{10-11} and \hkl{10-12} twin systems are simultaneously of  type I and II because in these cases, both operations are equivalent.
On the other hand, the atomic structures obtained by type I or type II twinning are not the same for the \hkl{11-21} and \hkl{11-22} twin systems. 
Structures corresponding to both types have been relaxed with \ai and the EAM potential.
It was found that type II is unstable for \hkl{11-22} and relaxes to the type I twin structure in both \ai and EAM.  
For \hkl{11-21} twins however, both types are stable but the type I structure has a much lower energy, 144 mJ\,m$^{-2}$ (229 mJ\,m$^{-2}$ in \textit{ab initio}) compared to 610 mJ\,m$^{-2}$ for the type II structure (749  mJ\,m$^{-2}$ in \textit{ab initio}).
This is in agreement with previous \ai calculations, not only in Zr \cite{Kumar2015,Morris1994,Morris1995} but also in other hcp elements,\cite{Jong2015,Ni2015} and with the structures observed experimentally with high resolution electron microscopy in various hcp metals such as Ti,\cite{Kasukabe1993, Pond1995, Braisaz1996} Zn,\cite{Lay1994, Braisaz1997} Co,\cite{Zhang2012} or Mg.\cite{Sun2014}
The structures predicted by the EAM potential (figure \ref{fig:Structures_EAM}) agree with \ai results with a slight exception for the \hkl{11-21} twin boundary (figure \ref{fig:Structures_EAM}d).
The stable structure obtained with both the EAM potential and by \ai is of type I.
With the EAM potential however, there is an additional translation parallel to the interface.
The amplitude of this displacement, 0.07 $a$ is small ($a=3.234$ \AA{} is the lattice parameter). 
The interface without the translation (structure drawn in grey in fig. \ref{fig:Structures_EAM}d) has an energy only 3 mJ\,m$^{-2}$ higher than the fully relaxed structure but is unstable.
This artifact of the EAM potential is  therefore considered harmless for the study of the \hkl{11-21} twinning system.
A similar stable structure for this perfect twin boundary has been observed by Bacon and Serra \cite{Bacon1991} in Ti with a many-body potential of the Finnis-Sinclair form.

\subsection{Energies of twin boundaries} 

\begin{table}[!tb]
	\caption{Perfect twin boundaries energies in mJ\,m$^{-2}$.
	The regime of strain corresponding to the experimental activation of the twinning system \cite{Yoo1991}
	is mentioned by C (compression) or T (tension) and the temperature regime by HT (high temperature) and LT (low temperature).}
	\begin{ruledtabular}
	\begin{tabular}{lllll}
	&        &  Twin  &  EAM  & GGA\\
	\hline \vspace{-10pt}\\	
	C& HT & \hkl{10-11}& 150 & 96  \\
	 & LT & \hkl{11-22}& 209 & 355 \\
	T& HT & \hkl{10-12}& 264 & 272 \\
	 & LT & \hkl{11-21}& 144 & 229 \\
	\end{tabular}
	\label{tab:Perfect_twin_energies}
	\end{ruledtabular}
\end{table}

The energies of perfect twin boundaries are gathered in table \ref{tab:Perfect_twin_energies}.
Different values are obtained with the EAM potential and the \ai calculations.
However, if one looks separately to the twin systems activated under compression 
or tension, \cite{Yoo1991} the same relative stability is obtained with both methods. 
For compression twins, the \hkl{10-11} boundary is more stable than \hkl{11-22},
and for tension, \hkl{11-21} is more stable than \hkl{10-12}. 
A qualitative agreement is obtained between the EAM potential and \ai calculations,
similar to the agreement previously obtained on stacking fault energies.\cite{Mendelev2007,Chaari2014}

One can conclude that this empirical potential leads to a reasonable description of the various twinning systems that can be activated in Zr, thus justifying its use to study twin growth at an atomic scale.

\section{Disconnection formation}
\label{sec:Formation}

Twin boundaries move thanks to the migration of disconnections.\cite{Christian1995,Hirth2016}
Disconnections are formed either by absorption of dislocations coming from the bulk \cite{Serra1995, Fan2015, Fan2016, Wang2012} or by nucleation on the twin boundary.\cite{Luque2014}
Once formed, disconnections migrate along the twin boundary and propagate the twinned crystal.
The following section describes the structure of the different disconnections that can appear on the four twinning systems activated in Zr and define their formation energy, before addressing their migration in the next section. 

\subsection{Crystallography}

\begin{figure}[!b]
		\subfigure[\hkl{10-11}]{
		\includegraphics[width=.46\linewidth]{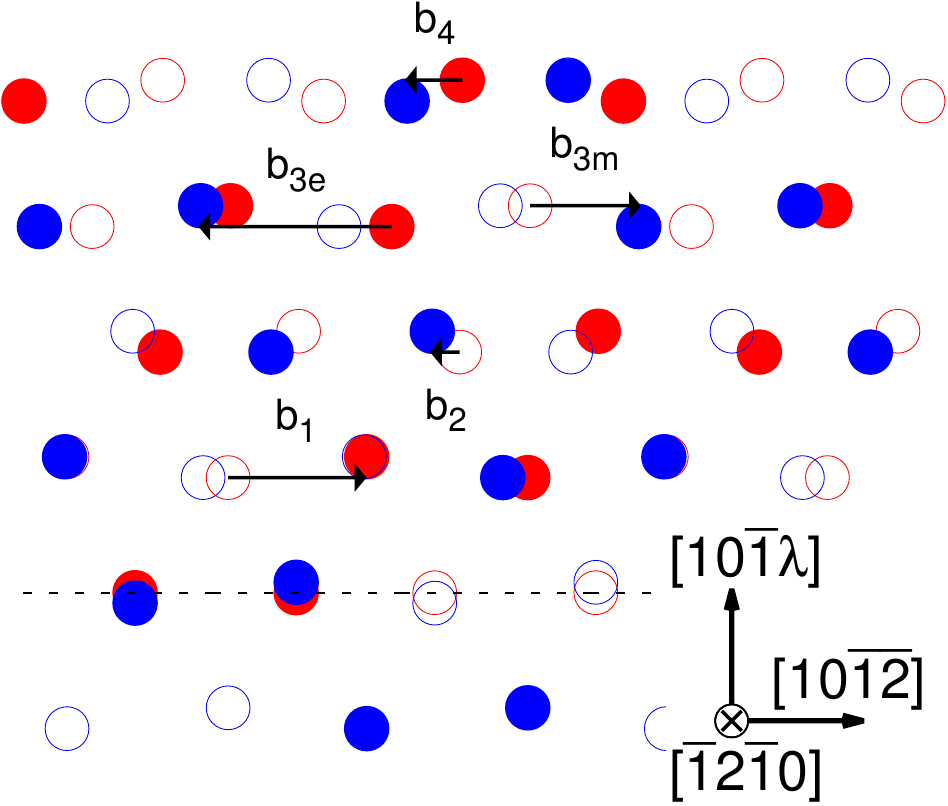}
		}
		\hfill \subfigure[\hkl{11-22}]{
		\includegraphics[width=.46\linewidth]{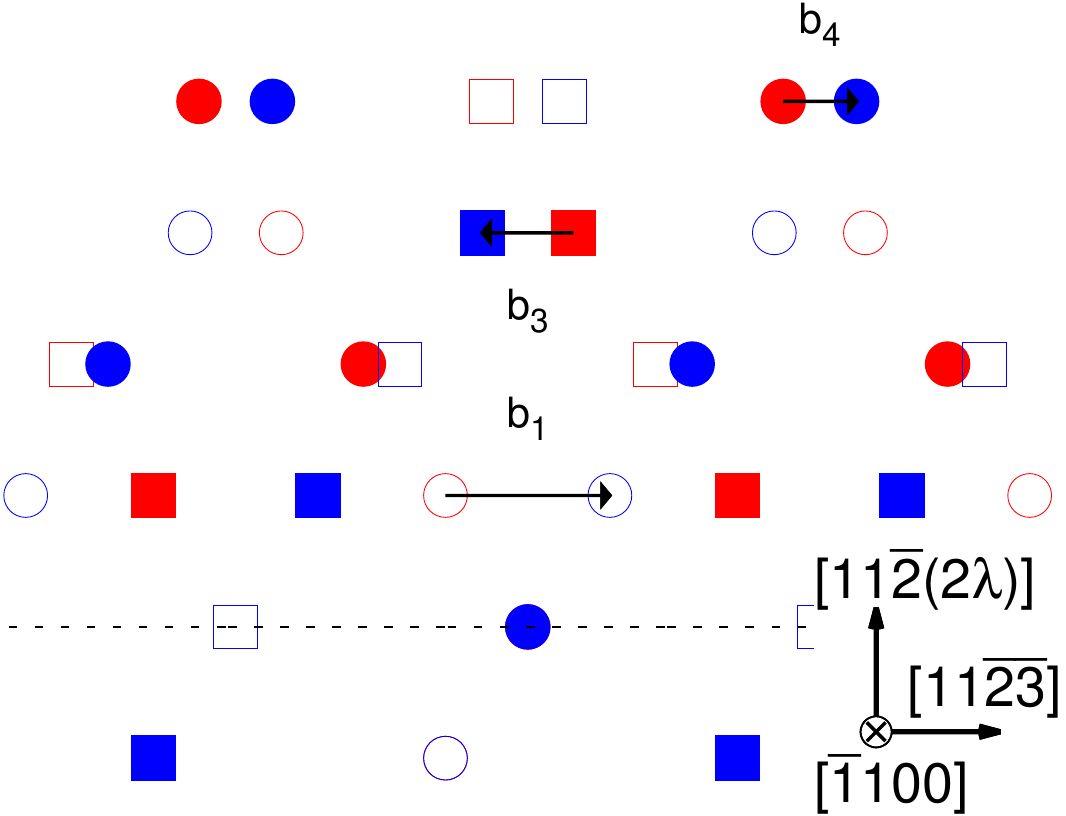}
		}\\
		\subfigure[\hkl{10-12}]{
		\includegraphics[width=.46\linewidth]{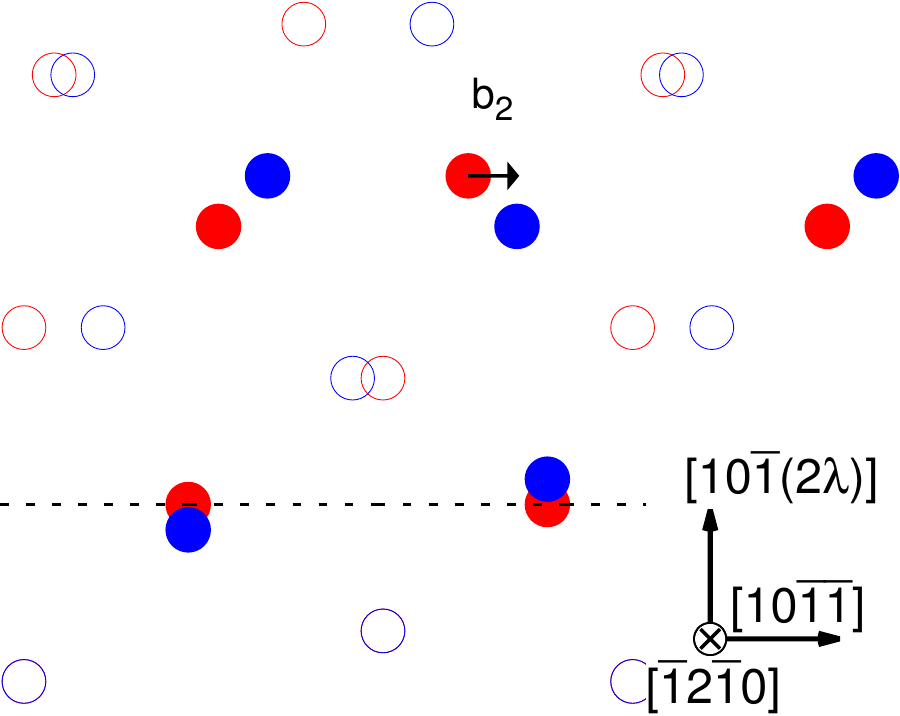}
		}
		\hfill \subfigure[\hkl{11-21}]{
		\includegraphics[width=.46\linewidth]{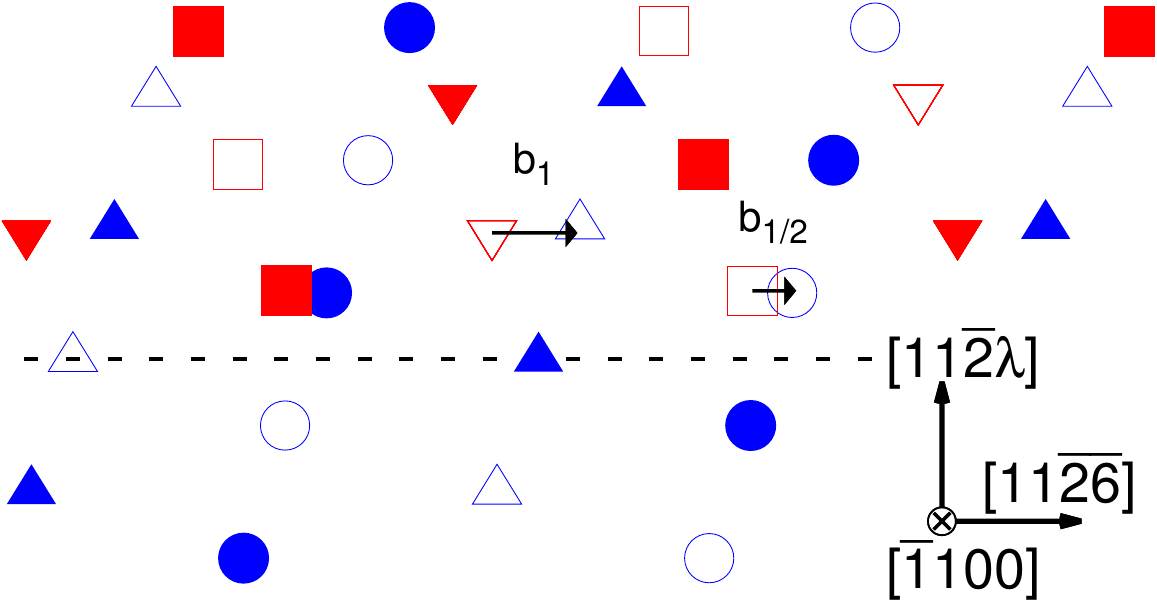}
		}
	\caption{Dichromatic patterns for the four twin systems
	showing the different possible Burgers vectors \burg{h}.
	}
	\label{fig:Dichromatic_diagrams}
\end{figure}

Disconnections are defined as steps along the twin boundary with a dislocation character.\cite{Hirth1996}
The step height is called $h$ and for purpose of simplicity is hereafter divided by the interplanar spacing $d_{K_1}$, so that $h$ is an integer. 
The dislocation content is characterized by the Burgers vector \burg{h} corresponding to the shear needed to make the twin grow by $h$ planes at the expense of the parent crystal.
Several disconnections corresponding to different $h$ and \burg{h} can exist for a given twinning system.  
Keeping only the possibilities leading to the smallest $h$ and \burg{h}, as in previous atomistic studies,
\cite{Serra1988,Serra1991,Serra1995,Serra1996,Wang2011,Li2012,Khater2013}
table \ref{tab:E_and_k} gathers the crystallographic definitions of the different disconnections considered in this study
(see also the dichromatic patterns in figure \ref{fig:Dichromatic_diagrams}).
All these disconnections have been found stable.
In this table, the Burgers vector is decomposed into an edge and a screw component, with the line direction $\vec{\zeta}$ of the disconnection taken normal to the experimental twinning direction $\vec{\eta}_1$.\cite{Christian1995}
Due to this choice, the twin modes observed experimentally correspond to pure edge disconnections.
The  orientation has also been chosen so that compression (respectively tension) along the \hkl<c> axis activates twinning through the glide
of disconnections with negative (respectively positive) edge component.

\begingroup
\begin{table*}[h!b!t]
	\caption{
	Crystallographic definition of the disconnections found stable in zirconium for different twinning systems.
	The four systems are defined by their twin plane $\mrm{K_1}$,
	their experimental twinning vector $\vec{\eta}_1$,
	and their vector $\vec{\zeta}$ normal to the experimental shear plane. $d_{K_1}$ is the interplanar distance between two $\mrm{K_1}$ planes. 
	The disconnection Burgers vectors \burg{h} are decomposed into edge and screw components, $b_e$ and $b_s$, respectively along $\vec{\eta}_1$  and $\vec{\zeta}$.
	The sign of the edge component is calculated for Zr ($\gamma$=1.598).
	$\gamma$ is the axial $c/a$ ratio and every distance in this table is normalized by the lattice constant $a$.
	Energy properties of the disconnections are also defined: 
	core energy $E^{\rm core}$, elastic energy prefactor $k$, core radius $r_{\rm c}$ entering the definition of the formation energy (Eq. \ref{eqn:E_form_final_form}),
	migration energy $E^{\rm mig}$, and Peierls stress $\tau_{\rm P}$ 
       }
	\begin{ruledtabular}
	\begin{tabular}{ccccccccc}
	\burg{h} & $b_e$ & $b_s$ & ||${\vec{b}_{h}}$|| & $ E^\mrm{core}$ & $k$ & $r_{\rm c}/h$ & $E^\mrm{mig}$ & $\tau_{\rm P}$ \\
		 	 &      &       &                     & \scriptsize{(meV\,\AA$^{-1}$)} & \scriptsize{(meV\,\AA$^{-3}$)}&  & \scriptsize{(meV\,\AA$^{-1}$)} & \scriptsize{(GPa)}\\
	\hline \vspace{-6pt}	 \\
	\multicolumn{9}{c}{$\mrm{K_1}$=\hkl{10-11}; $d_{K_1}$=$\dfrac{\sqrt3\gamma}{\sqrt{4\gamma^2+3}}$; $\vec{\eta}_1$=$\lbrack 10\bar{1}\bar{2} \rbrack $; $\vec{\zeta}$=$\lbrack1\bar{2}10\rbrack $} \\
		\burg{1} & $\dfrac{3}{\sqrt{4\gamma^2+3}} >0$  			& 0               			&0.826	& -87	& 742 & 0.5	& 10 		& 0.30  \\	
		\burg{2} & $\dfrac{9-4\gamma^2}{2\sqrt{4\gamma^2+3}} <0$ 	& $\pm\dfrac{1}{2}$ 			&0.529	& -4 	& 544 & 0.28	& 6		& 0.17  \\	
		\burger{3e}{-/+}& $\dfrac{2(3-2\gamma^2)}{\sqrt{4\gamma^2+3}} <0$& 0			&1.08	&  236	& 742 & 0.5	& 120 / 208	& 0.21 / $> 0.21$ \\	
		\burger{3m}{-/+}& $\dfrac{15-4\gamma^2}{2\sqrt{4\gamma^2+3}} >0$	& $\pm\dfrac{1}{2}$	&0.826	&  289	& 648 & 0.37	& 25 / 20 	& $> 0.39$ /  0.39 \\	
	\vspace{3pt}	\burg{4} & $\dfrac{9-4\gamma^2}{\sqrt{4\gamma^2+3}} <0$	& 0 					&0.334	&  117	& 742 & 0.5	& 116		& $> 2$  \\	
	\hline \vspace{-6pt}\\
 	\multicolumn{9}{c}{$\mrm{K_1}$=\hkl{11-22}; $d_{K_1}$=$\dfrac{\gamma}{2\sqrt{\gamma^2+1}}$; $\vec{\eta}_1$=$\lbrack 11\bar{2}\bar{3} \rbrack $; $\vec{\zeta}$=$\lbrack1\bar{1}00\rbrack $ }\\
		\burg{1}  	&$\dfrac{1}{\sqrt{\gamma^2+1}} >0$  	 	& 0 &0.529	& -74	& 750 & 0.5	& 0.2 &	0.02 	  \\ 	
		\burger{3}{}	&$\dfrac{2-\gamma^2}{\sqrt{\gamma^2+1}} <0$	& 0 &0.294	&  119  & 750 & 0.5	& 22  & 1.87 	  \\	
	 	\burger{3}{-/+} &$\dfrac{2-\gamma^2}{\sqrt{\gamma^2+1}} <0$ & 0 	&0.294	&  237  & 750 & 0.5	& 40 / 10  & $> 1.47$ / 1.47  \\	
		\burger{4}{}	&$\dfrac{3-\gamma^2}{\sqrt{\gamma^2+1}} >0$ & 0 &0.238	&  254  & 750 & 0.5	& 32  & 1.82	  \\	
	\vspace{3pt}	\burger{4}{-/+}	&$\dfrac{3-\gamma^2}{\sqrt{\gamma^2+1}} >0$	& 0 &0.238	&  250  & 750 & 0.5	& 22 / 14 & 2.38 / $> 2.38$ \\	
	\hline \vspace{-6pt}\\
	\multicolumn{9}{c}{$\mrm{K_1}$=\hkl{10-12}; $d_{K_1}$=$\dfrac{\sqrt3\gamma}{2\sqrt{3+\gamma^2}}$; $\vec{\eta}_1$=$\lbrack 10\bar{1}\bar{1} \rbrack $; $\vec{\zeta}$=$\lbrack1\bar{2}10\rbrack $}  \\
	\vspace{3pt}	\burg{2} 	&$\dfrac{3-\gamma^2}{\sqrt{3+\gamma^2}} >0$ & 0 & 0.189	&   5	& 753 & 0.5 	& 2	& 0.15	  \\	
	\hline  \vspace{-6pt}\\
	\multicolumn{9}{c}{$\mrm{K_1}$=\hkl{11-21}; $d_{K_1}$=$\dfrac{\gamma}{\sqrt{4\gamma^2+1}} $; $\vec{\eta}_1$=$\lbrack 11\bar{2}\bar{6} \rbrack $; $\vec{\zeta}$=$\lbrack1\bar{1}00\rbrack $}  \\
	\vspace{3pt}	\burg{1/2} 	& $\dfrac{1}{2\sqrt{4\gamma^2+1}} >0$	& 0 & 0.149 	& -25	& 758 & 0.5 	& < 0.1 & < 0.01 \\ 
	\end{tabular}%
\end{ruledtabular}	
	\label{tab:E_and_k}
\end{table*}
\endgroup

From a purely elastic point of view, the disconnections with the smallest Burgers vectors are expected to have the lowest formation energies.
However, the steps also induce atomic rearrangements that represent a core energy cost. 
This contribution can be intuitively expected to be minimal for the smallest step heights.
Since the smallest Burgers vectors are usually not obtained for the smallest step heights (Tab. II), the competition between the core and elastic energies prevents to predict which disconnection is the most stable for a given twinning system.

In the particular case of the \hkl{11-21} twin system, the \burg{1} disconnection  is not stable
and spontaneously dissociates into two \burg{1/2} disconnections.\cite{Serra1988,Serra1991,Khater2013}
Although the height of this disconnection is not an integer, the \burg{h} notation will be kept.

A shuffling is imposed to the atoms between the former and the new twin planes following the method described by Serra \textit{et al}.\cite{Serra1988,Serra1991}
Several initial shuffling may be possible, leading in some cases to different disconnection cores after relaxation, that is to say disconnections with the same \burg{h} and $h$ but different atomic structures and thus different formation and migration energies.  
This happens for \burg{3} and \burg{4} disconnections on \hkl{11-22} twin planes,
where two different configurations of the disconnection dipole can be stabilized,
depending on the initial shuffling.

Most disconnection dipoles are symmetrical: both disconnections are then equivalent.
But some asymmetric dipoles have also been found. 
The two different core regions are then characterized by a different atomic density.
The disconnection with the higher (respectively the lower) density will be labeled \burger{h}{+} (respectively \burger{h}{-}).

\subsection{Atomistic simulations}

\begin{figure}[!b]
	\begin{center}
	\includegraphics[width=.45\textwidth]{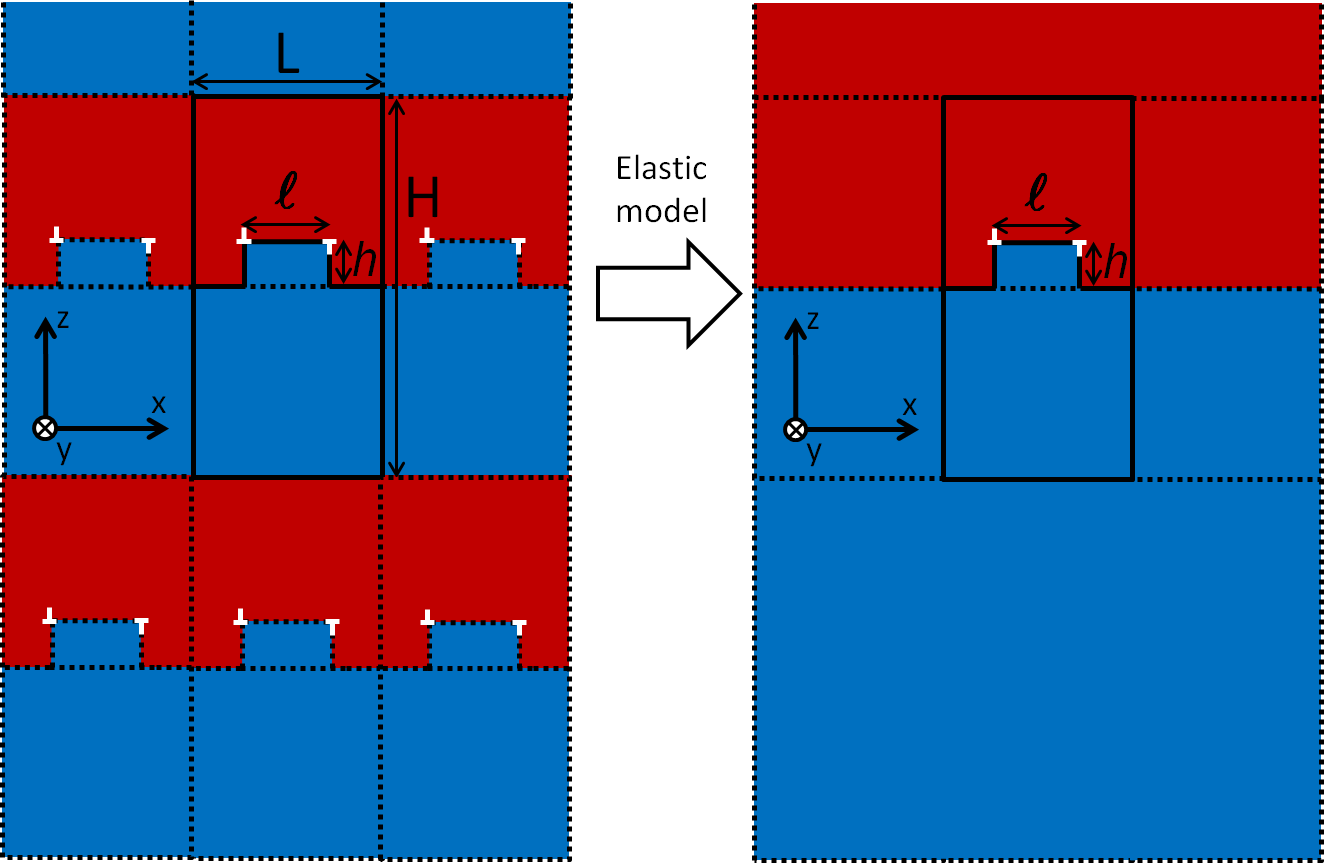}
	\end{center}
	\caption{Sketch of the set-up used in atomistic simulations with full periodic boundary conditions (left) 
	and corresponding isolated dipole configuration (right) deduced through elastic modeling.}
	\label{fig:Ideal_problem}
\end{figure}
 
The atomistic simulations of disconnections are performed with the EAM potential described in the previous section.
The twin plane lies in the $(xy)$ plane with the $x$-direction along the $\vec{\eta}_1$ twinning direction.
The unitary cell is repeated from thirty up to two hundred times in this direction, leading to a cell length $L$
along $x$ (Fig. \ref{fig:Ideal_problem}).
The cell dimension $H$ in the $z$-direction normal to the twin plane is chosen large enough for the twin boundary not to interact with its periodic images.
In practice, cells with up to sixty planes between the two twin boundaries have been used.

\begin{figure}[!bt]
	\begin{center}
	\includegraphics[width=.45\textwidth]{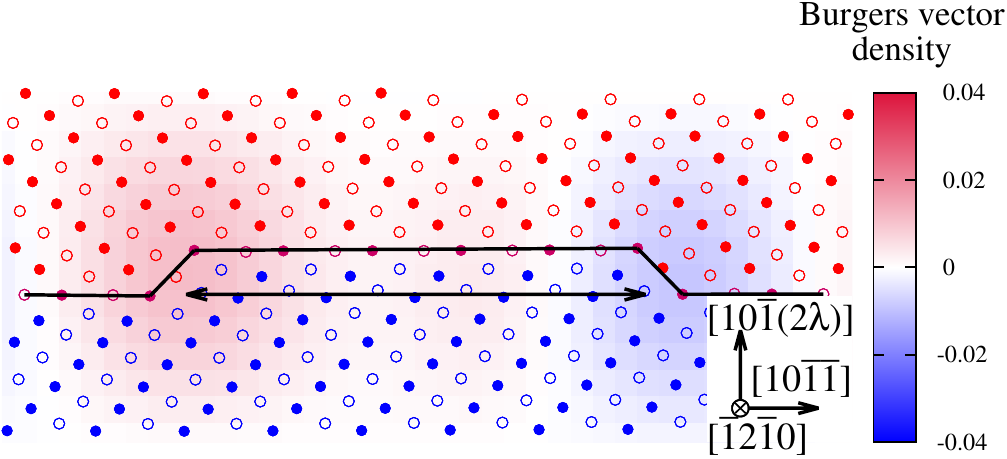}
	\end{center}
	\caption{Relaxed structure of the \burg{2} disconnection dipole along the \hkl{10-12} twin boundary. 
	Colored areas represent the Burgers vector density (normalized by the lattice parameter $a$) corresponding to the Nye tensor. \cite{Hartley2005} The double headed arrow shows the length $\ell$ of the disconnection dipole deduced through equation \ref{eqn:boxes_strain} from the total plastic strain.} 
	\label{fig:Nye_tensor_10_12}
\end{figure}

A disconnection dipole of length $\ell$, with a line vector $\vec{\zeta}$ along the $y$-direction,
is introduced in the simulation box 
by displacing all atoms according to the elastic field created by a dislocation dipole 
of Burgers vector \burg{h} located at a distance $h$ above the twin boundary,
taking full account of periodic boundary conditions.\cite{Cai2003}

An homogeneous strain $\varepsilon^0$, corresponding to the plastic strain introduced in the simulation cell 
by the creation of the disconnection dipole, is also applied to cancel the stress in the cell.\cite{Rodney2017} 
Its non-zero components are given by:
\begin{equation}
	\varepsilon^0_{k3} = \varepsilon^0_{3k} = \dfrac{b_k \ell}{2HL}	.
	\label{eqn:boxes_strain}
\end{equation}
Atomic positions are then relaxed thanks to a conjugate gradient algorithm.
As this relaxation can change the length $\ell$ of the dipole, a residual stress may be observed after relaxation. 
The applied homogeneous strain is adjusted to cancel this stress.
Further relaxation does not change the atomic structure of the disconnections and leads to zero stress.
In the following, the actual length $\ell$ of the disconnection dipole will be defined through equation \ref{eqn:boxes_strain},
using the strain $\varepsilon^0_{13}$ that has been applied to the simulation box to cancel the stress.
In cases where the applied strain $\varepsilon^0_{23}$ is not zero, i.e. when the disconnection has a screw component,
it has been found that the lengths $\ell$ deduced either from $\varepsilon^0_{13}$ or $\varepsilon^0_{23}$ do not differ by more than 1 \AA.

Figure \ref{fig:Nye_tensor_10_12} represents the relaxed atomic structure of a \burg{2} disconnection dipole on the \hkl{10-12} twin boundary.
Two steps can be clearly seen on the interface separating the parent and twinned crystals.
Away from these steps, one recovers the structure of the perfect twin boundary.
Nye tensor can be extracted from the relaxed configurations \cite{Hartley2005} to visualize the Burgers vector density created by both disconnections. 
The density is located at the steps and vanishes over a distance approximately equal to five times the lattice parameter $a$,
confirming the localized character of the shear incompatibility.
The dipole length $\ell$ calculated with equation \ref{eqn:boxes_strain} is reported on this figure.
This definition of $\ell$ agrees with the dipole length that can be inferred 
from the relaxed atomic structure.
The same behavior is observed for the different disconnection dipoles studied in the present work.

\subsection{Formation energy}
\label{subsec:Formation_energy_of_disconnection_dipoles}

\begin{figure}[!b]
	\includegraphics[width=.45\textwidth]{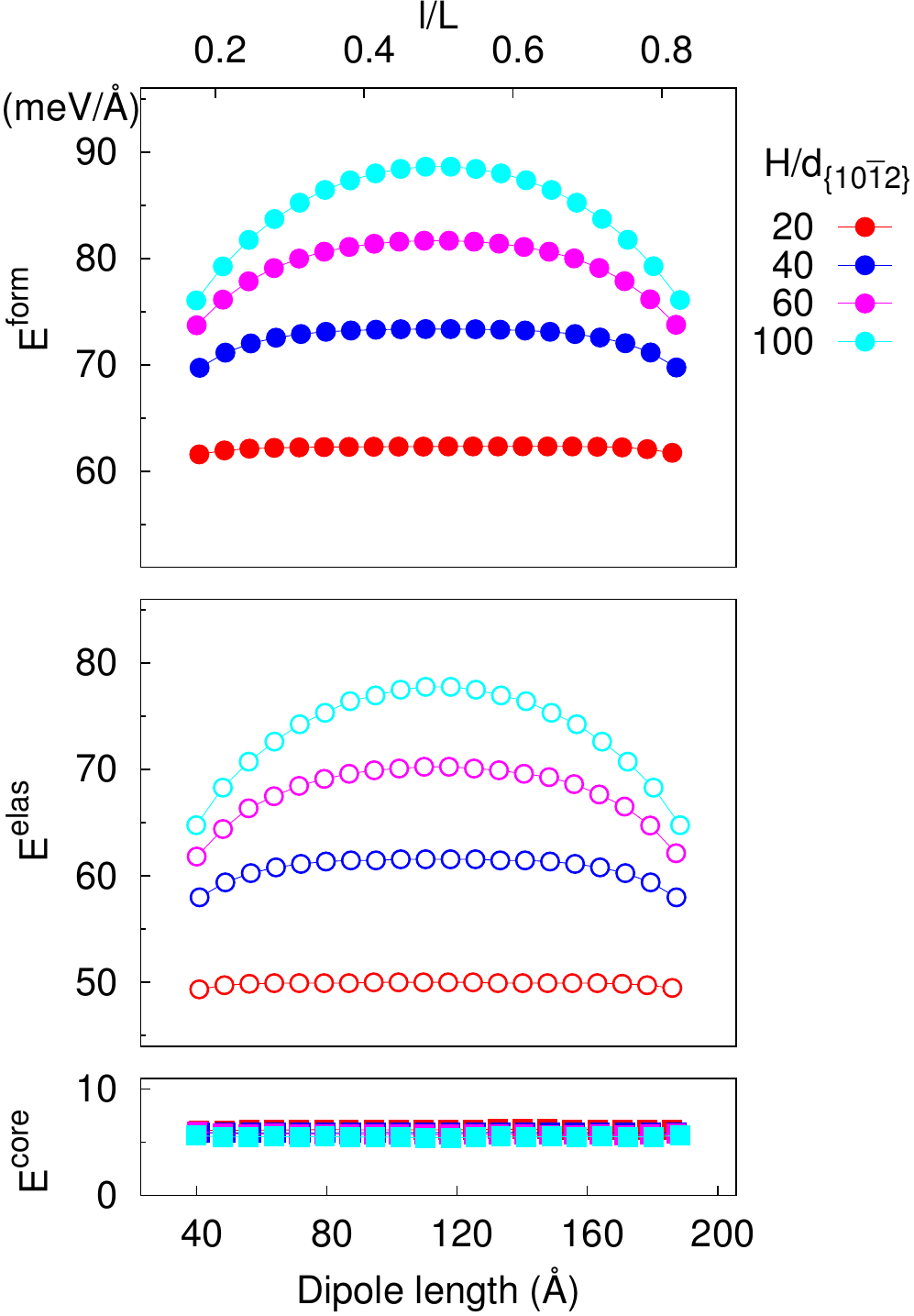}
	\caption{Formation energy $E^{\rm form}$, and its decomposition into an elastic $E^{\rm elas}$ 
	and an atomic $E^{\rm core}$ contributions, 
	of the \burg{2} disconnection dipole on the \hkl{10-12} twin boundary
	as a function of the dipole length $\ell$ for four different simulation cell heights $H$.
	The width of the simulation cell is $L=228$\,\AA.
	}
	\label{fig:Variations_hauteur}
\end{figure}

The formation energy $E^\mrm{form}$ of a disconnection dipole is defined as the energy difference
between a simulation cell containing a disconnection dipole along one of its twin boundary
and the same cell with two perfect twin boundaries, both cells being under zero stress.
The formation energy depends not only on the length $\ell$ of the disconnection dipole, but also on the dimensions of the simulation cell. 
This is illustrated in figure \ref{fig:Variations_hauteur} where the formation energy of the $\vec{b}_2$ disconnection dipole on the \hkl{10-12} twin boundary is displayed as a function of $\ell$ for various cell heights $H$ and a fixed cell width $L$. 
The energy variations are caused by the interactions between the two disconnections composing the dipole and their periodic images.
As these interactions are elastic, one can decompose the formation energy in two parts, an elastic and a core contribution:
\begin{equation}
E^\mrm{form}(\ell,H,L)=   E^\mrm{elas}(\ell,H,L) + 2E^\mrm{core}.
	\label{eqn:E_form}
\end{equation}
The elastic energy $E^\mrm{elas}$, which contains all the interactions with the surrounding microstructure,
varies with the cell dimensions and with the dipole length.
On the other hand, the core contribution $E^\mrm{core}$ accounts for the cost of atomic disorder at the very core of the disconnections, which cannot be described by linear elasticity.
This contribution should be an intrinsic property of the disconnections and thus should not depend on the surrounding microstructure.
For symmetric dipoles, both disconnections have the same core energy, explaining the factor 2 in equation \ref{eqn:E_form}.
For asymmetric dipoles, $E^{\rm core}$ is then the average core energy between the \burger{h}{+} and \burger{h}{-} disconnections.

\begin{figure}[tb]
	\begin{center}
	\includegraphics[width=.43\textwidth]{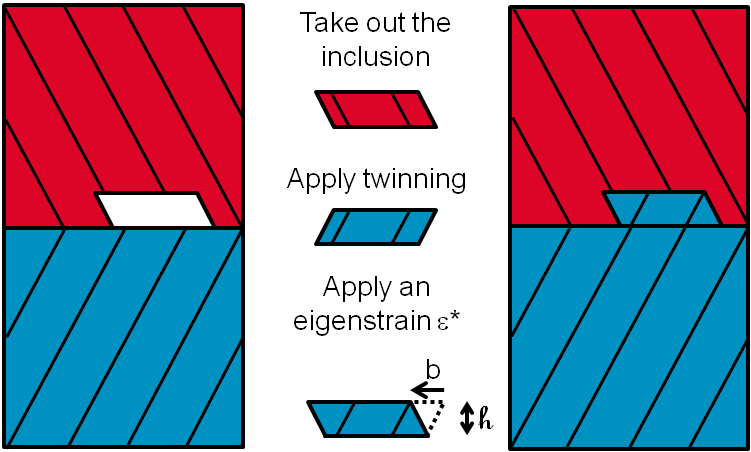}
	\end{center}
	\caption{Inclusion model used to calculate the elastic energy of a disconnection dipole.
	}
	\label{fig:Moulinec_Suquet}
\end{figure}

To check the validity of the energy decomposition in equation \ref{eqn:E_form}, the elastic energy of the disconnection dipole is calculated separately. 
In this elastic calculation, the dipole is modeled as an Eshelby inclusion \cite{Mura1987} of length $\ell$ and height $h$ with an eigenstrain $\varepsilon^*_{i3} = b_i/2h$ (Fig. \ref{fig:Moulinec_Suquet}).
The elastic energy is computed within linear elasticity theory thanks to the Fast Fourier Transform (FFT) approach of Moulinec and Suquet,\cite{Moulinec1998} taking full account of elastic anisotropy, elastic inhomogeneity, and periodic boundary conditions.

As one can see in figure \ref{fig:Variations_hauteur}, the calculated elastic energy shows the same variations with dipole length and cell dimensions as the formation energy extracted from the atomistic simulations, confirming that these variations are elastic in nature.  
Withdrawing the elastic contribution from the formation energy, one obtains a constant energy contribution corresponding to the core energy (Fig. \ref{fig:Variations_hauteur}).
The variation of core energies obtained with such a procedure does not exceed 2 meV \AA$^{-1}$, confirming that the decomposition of the formation energy into elastic and core contributions (Eq. \ref{eqn:E_form}) is meaningful. 

The core energies obtained for all disconnection dipoles are given in table \ref{tab:E_and_k}.
As the elastic energy may be larger than the formation energy, negative core energies are obtained for some of the disconnections.  
We will show in the next paragraph that this is simply a consequence of the energy decomposition assumed when modeling the disconnection dipole as an Eshelby inclusion.
One has to remember that this decomposition is arbitrary and that despite those negative core energies, the formation energy of all dipoles is positive.

\subsection{Isolated disconnection dipoles}

Now that the decomposition of the formation energy into an elastic 
and a core contribution (Eq. \ref{eqn:E_form}) is validated, the elastic energy of an isolated dipole is calculated in order to withdraw the interactions with the periodic images. 

\begin{figure}[!b]
	\includegraphics[width=.98\linewidth]{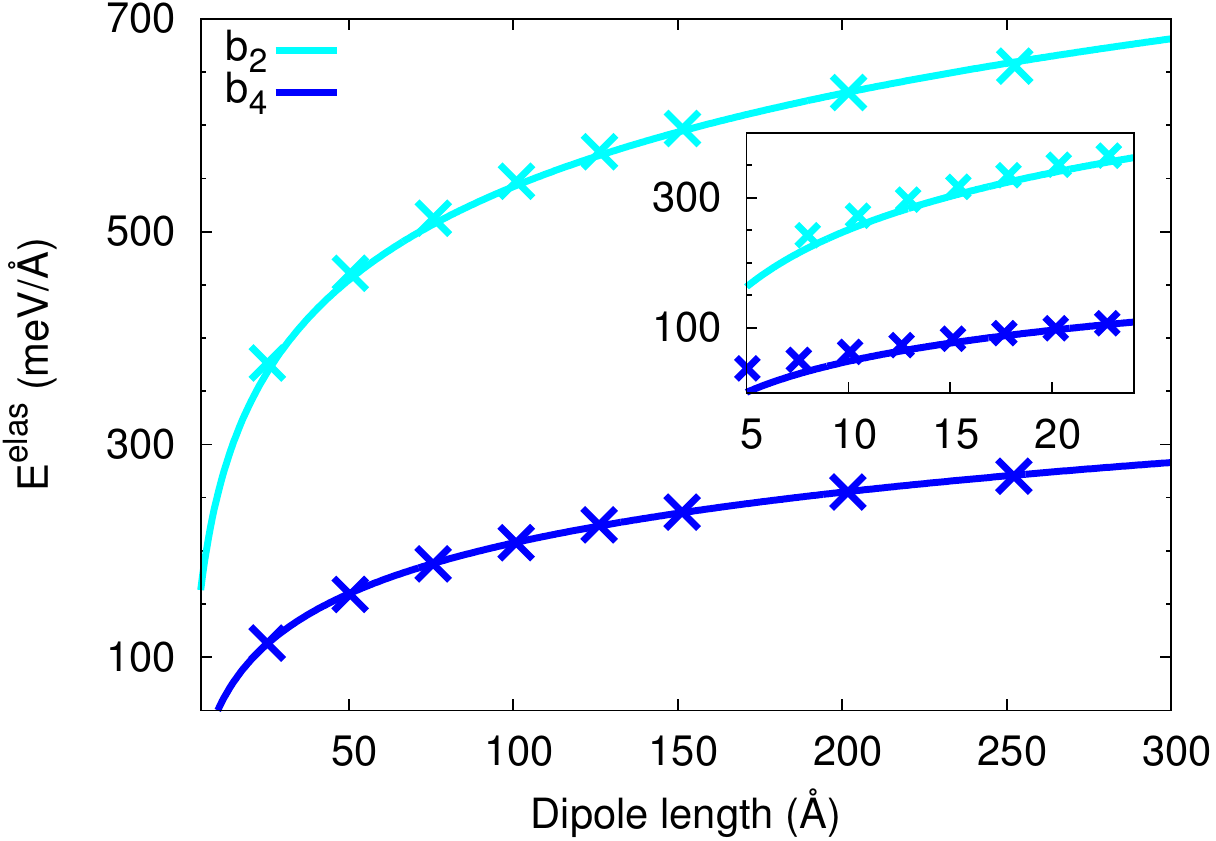}
	\caption{Comparison of the elastic energy obtained either with the FFT method (crosses) 
	or with the analytical expression (line) given in equation \ref{eqn:E_elas}
	for \burger{2}{}  and \burger{4}{} disconnection on the \hkl{10-11} twin plane.
	The inset highlights the differences observed for the lowest dipoles lengths.
	The FFT calculations are performed with $L=H=1260$\,\AA.}
	\label{fig:Isolated_elasticity}
\end{figure}

The elastic energy of an isolated dipole is calculated using the same FFT approach as in the previous section.
In practice, the cell used for this calculation remains periodic, 
but the cell dimensions $H$ and $L$ are chosen much larger than the dipole dimensions $h$ and $\ell$,
so that the result becomes independent of these cell dimensions. 
Figure \ref{fig:Isolated_elasticity} shows that the elastic energy varies as the logarithm of the dipole length following the same expression as for a dislocation dipole:
\begin{equation}
	E^\mrm{elas}(\ell)= \dfrac{1}{4\pi}kb^2\ln \left( \dfrac{\ell}{r_{\rm c}} \right).
	\label{eqn:E_elas}
\end{equation}
The FFT calculations differ from this analytical expression only at small dipole lengths.
However, as soon as the dipole length becomes larger than its height ($\ell \gg h$), the elastic energy can be expressed analytically through Eq. \ref{eqn:E_elas}.
$k$ is a constant that depends only on the elastic constants and the disconnection orientation \cite{Dupeux1980} while $r_{\rm c}$ is an effective core radius.
These two parameters are obtained by fitting equation \ref{eqn:E_elas} to the results of the FFT calculations at the large dipole lengths and results are displayed in table \ref{tab:E_and_k}.
The main variations of $k$ are caused by the disconnection character, with $k$ being larger for pure edge than for mixed disconnections.
Elastic anisotropy also leads to a slight variation of this parameter.
On the other hand, the ratio $r_{\rm c}/h$ only depends on the disconnection character, with $r_{\rm c}\approx h/2$ for pure edge and smaller values for mixed orientations. 
The effective core radius entering the definition of the elastic energy (Eq. \ref{eqn:E_elas}) is therefore much smaller than the actual extension of the disconnection core, as can be inferred for instance from the spreading of the Burgers vector density obtained from the Nye tensor (Fig. \ref{fig:Nye_tensor_10_12}).
As a consequence, the region where linear elasticity breaks down is more extended than the cylinder of radius $r_{\rm c}$ around the disconnection line. 

Combining equations \ref{eqn:E_form} and \ref{eqn:E_elas} gives the formation energy of an isolated disconnection dipole. 
\begin{equation}
	E^\mrm{form}(\ell)= \dfrac{1}{4\pi}kb^2\ln \left( \dfrac{\ell}{r_{\rm c}} \right) + 2E^\mrm{core}.
	\label{eqn:E_form_final_form}
\end{equation}
The core radius inferred from FFT calculations being smaller than the actual extent of the disconnections is the reason why negative core energies are obtained for some of the disconnections (Tab. \ref{tab:E_and_k}).

\begin{figure}[!b]
		\subfigure{
		\includegraphics[width=.98\linewidth]{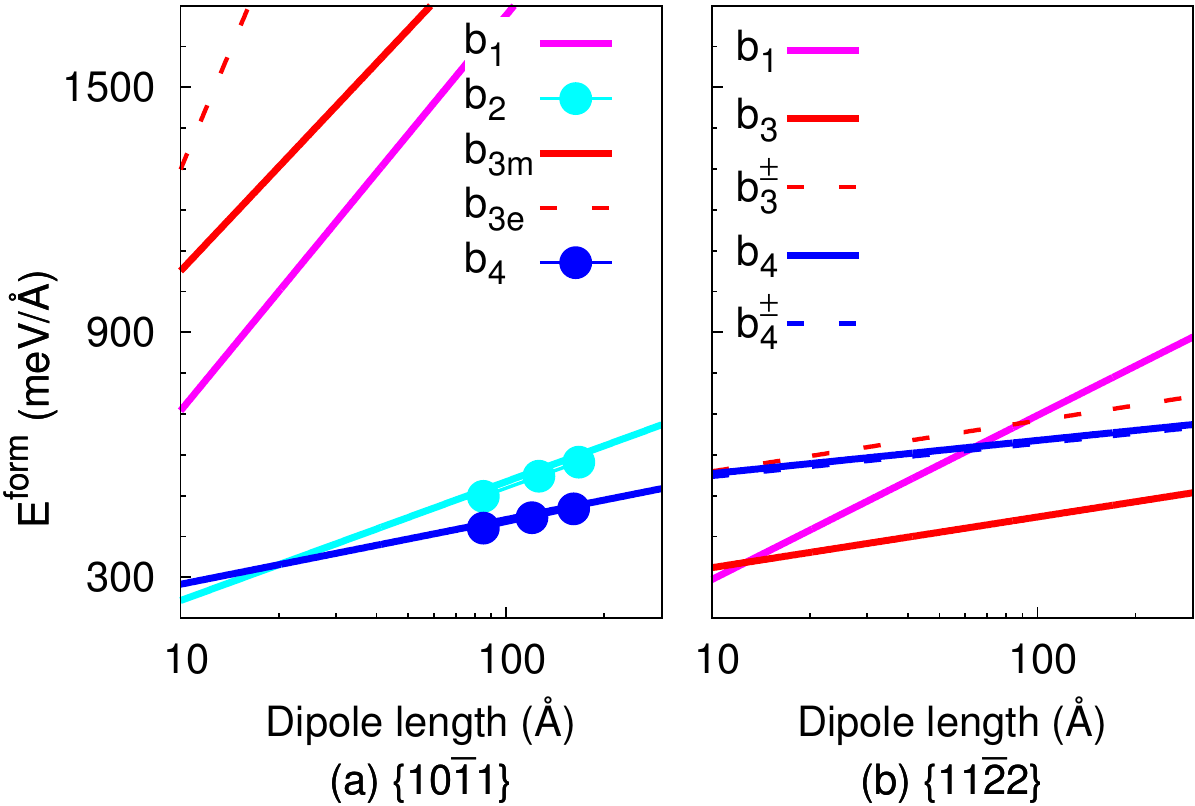}
		}
		\subfigure{
		\includegraphics[width=.98\linewidth]{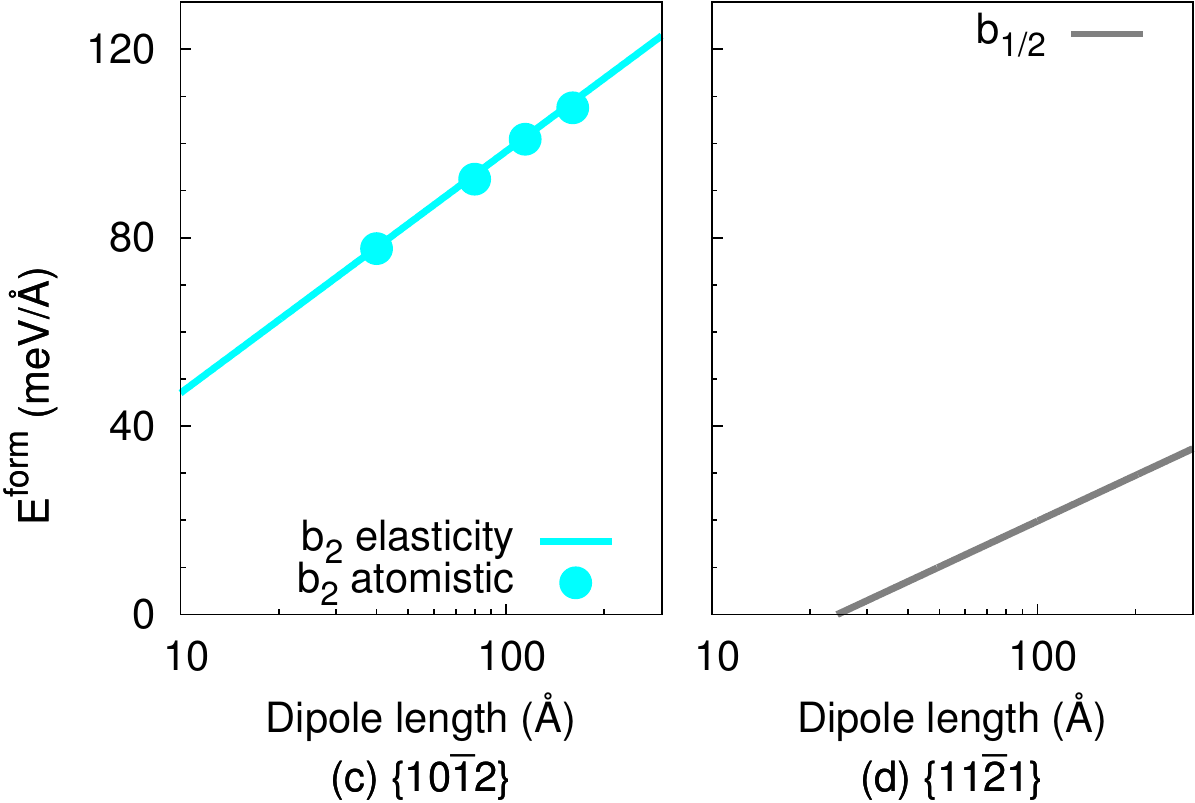}
		}
	\caption{Formation energy of an isolated disconnection dipole as a function of the dipole length $\ell$
	for the different disconnections that can exist on the four twin boundaries.
	A logarithmic scale is used for the abscissa.
	The symbols in subfigure a and c show the results of atomistic simulations whereas the lines correspond to the analytical expression (Eq. \ref{eqn:E_form_final_form}).
	Atomistic simulations are performed with (a) $L=2350$\,{\AA} and $H = 1575$\,{\AA};
	(b) $L=3048$\,{\AA} and $H=1519$\,{\AA}.}
	\label{fig:Isolated_dipoles}
\end{figure}

Figure \ref{fig:Isolated_dipoles} shows the formation energy (Eq. \ref{eqn:E_form_final_form}) as a function of the disconnection dipole length for all isolated disconnection dipoles that exist in zirconium. 
Since the atomic simulations are relying on a EAM potential, one can also use these simulations with large enough cells to obtain a disconnection dipole isolated from its periodic images.
This offers a way to validate the analytical model.
The points on figures \ref{fig:Isolated_dipoles}a and  \ref{fig:Isolated_dipoles}c are the results of those atomistic calculations and show a quantitative agreement with the analytical expression.
As the size of the simulation cells used to obtain the core energy are compatible with \ai calculations, one can conceive using the same approach based on the partition of the disconnection energy into a core and an elastic contributions to obtain an \ai prediction of the formation energy of the disconnection dipole.
This is however left for future work.

Because of the two contributions entering the formation energy, the most stable disconnection on a twin boundary is not always the one with the smallest Burgers vector. 
This is the case for the \hkl{11-22} twin boundary where the symmetric configuration of the \burg{3} disconnection dipole has a low core energy that  compensates its elastic energy.
As a consequence, in the considered range of dipole lengths ($10\leq\ell\leq300$ \AA), this disconnection dipole has a lower formation energy than the \burg{4} dipole (Fig. \ref{fig:Isolated_dipoles}), 
despite its higher Burgers vector (Tab. 3).
The \burg{4} disconnection dipole will become more stable only for lengths $\ell$ larger than 1.5\,10$^6$ \AA, 
a length too large to be meaningful for disconnection nucleation.
Both \burg{3} and \burg{4} disconnections are pure edge with a Burgers vector collinear to the twinning direction \hkl<11-2-3>, which has been experimentally reported for this \hkl{11-22} twin boundary.\cite{Rapperport1959,Rapperport1960,Partridge1967,Paton1970,Yoo1981,Christian1995}
But because their Burgers vectors have edge components of opposites signs (Tab. \ref{tab:E_and_k}), the \burg{3} disconnection dipole  propagates the twin under compression of the \hkl<c> axis,
whereas the \burg{4} is activated under tension.  
The lower formation energy obtained for the \burg{3} disconnection dipole is therefore in agreement with the \hkl{11-22} twinning system being active in compression.\cite{Akhtar1973a,Yoo1981}
The shear magnitude $s=-b_3/3d_{\hkl{11-22}}=0.231$ induced by the glide of this disconnection also corresponds to the magnitude $s=0.225$ determined experimentally.\cite{Rapperport1959,Rapperport1960}

For the \hkl{10-11} twin boundary, the \burg{4} disconnection dipole is the most stable for dipole lengths larger than $20$ \AA{} (Fig. \ref{fig:Isolated_dipoles}).
This disconnection has a Burgers vector in agreement with the twinning direction experimentally reported, \cite{Paton1970,Yoo1981,Christian1995} with a sign corresponding to the \hkl{10-11} twin system active under compression, \cite{Akhtar1973a,Yoo1981} and a shear amplitude $s=-b_4/4d_{\hkl{10-11}}=0.109$ corresponding to the amplitude $s=0.104$ experimentally assessed in Zr.\cite{Jensen1972} 
But for small dipole lengths, the \burg{2} dipole becomes more stable because of its low core energy. 
This disconnection dipole, which will also be active under compression, is not pure edge and thus corresponds to a different twinning direction. 
As noted by Serra \etal \cite{Serra1991} this second \hkl{10-11} twinning mode has been proposed \cite{Christian1995} but not observed experimentally.

For the \hkl{11-21} twin, our model gives a negative formation energy below 25\,\AA. 
This is unphysical and illustrates the limits of our approach for this very low energy disconnection:
because of the large spreading of this disconnection, one cannot assume that the two disconnections composing the dipole
are interacting only through their elastic field for such small separation distances.

Finally, it is worth noting that the formation energies of the most stable disconnections vary 
strongly between the four different twin systems (Fig. \ref{fig:Isolated_dipoles}). 
These energies are higher for the two systems active under compression, \hkl{10-11} and \hkl{11-22}, 
and lower for the tension systems, \hkl{10-12} and \hkl{11-21}. 
This is mainly due to the low core energies of disconnections activated in tension.

\section{Disconnection migration}
\label{sec:Migration}

After modeling disconnection dipole formation, we now focus on their migration. 
Both the migration energy and Peierls stress of a disconnection are calculated, 
i.e. respectively the energy barrier to be overcome by the disconnection to glide without any applied stress
and the resolved shear stress which cancels this barrier, thus allowing the disconnection to glide without the help of thermal activation.

\subsection{Migration energy}

\begin{figure*}[!bt]
		\subfigure[\hkl{10-11}]{
		\includegraphics[width=.3\textwidth]{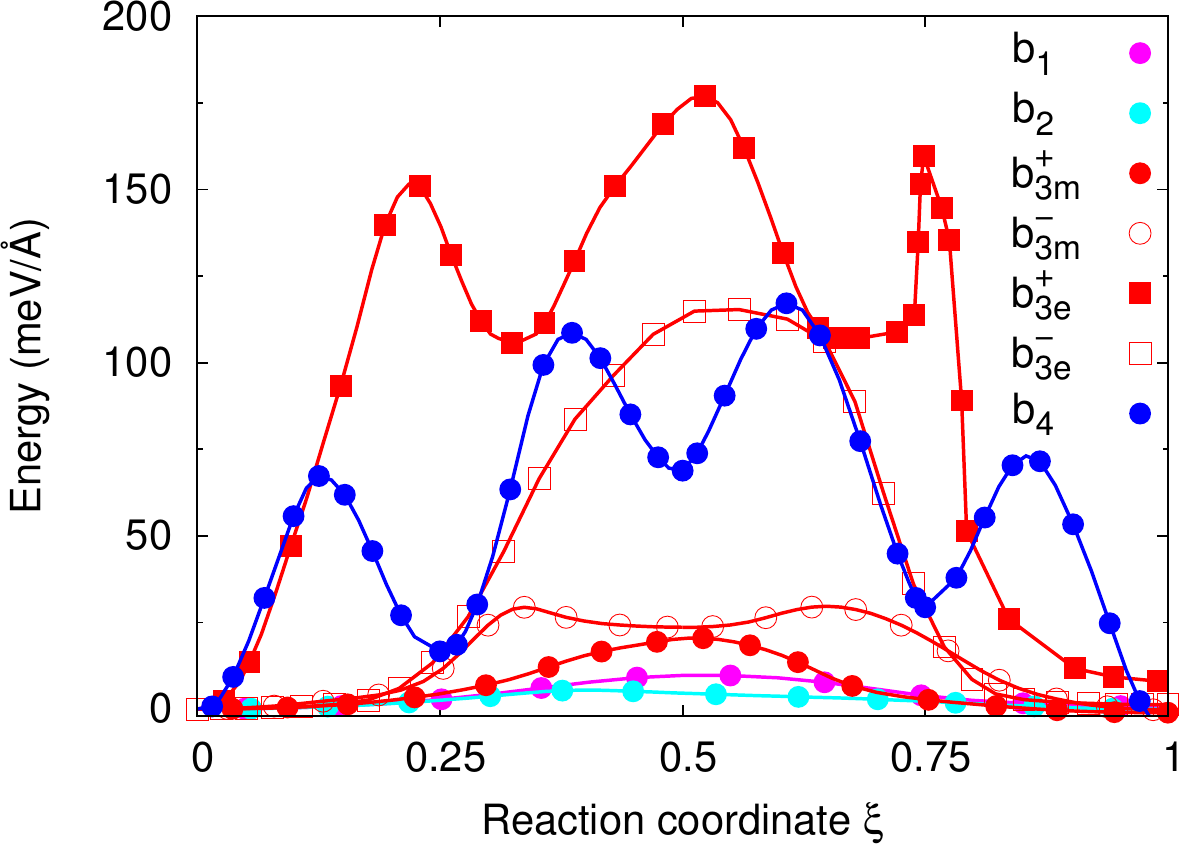}
		}
		\subfigure[\hkl{11-22}]{
		\includegraphics[width=.3\textwidth]{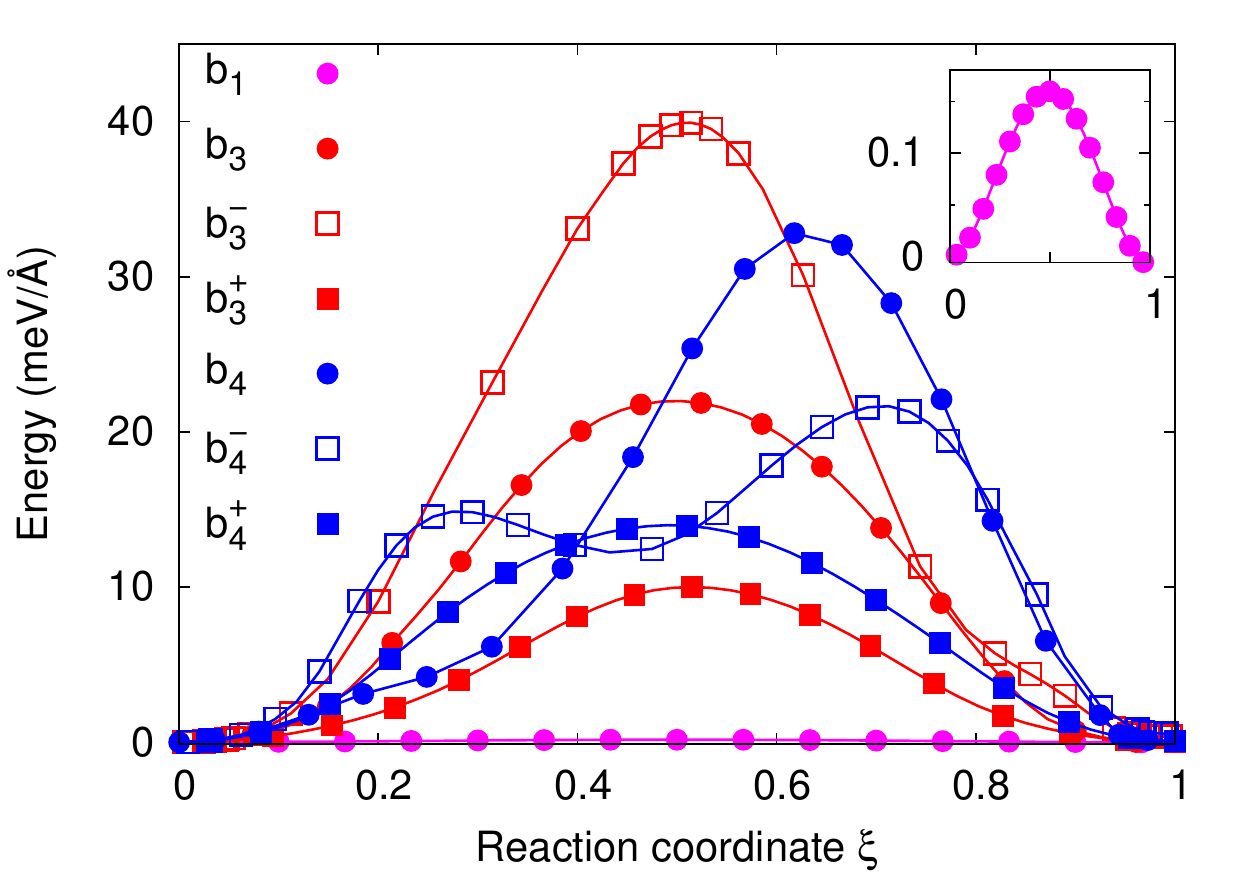}
		}
		\subfigure[\hkl{10-12}]{
		\includegraphics[width=.3\textwidth]{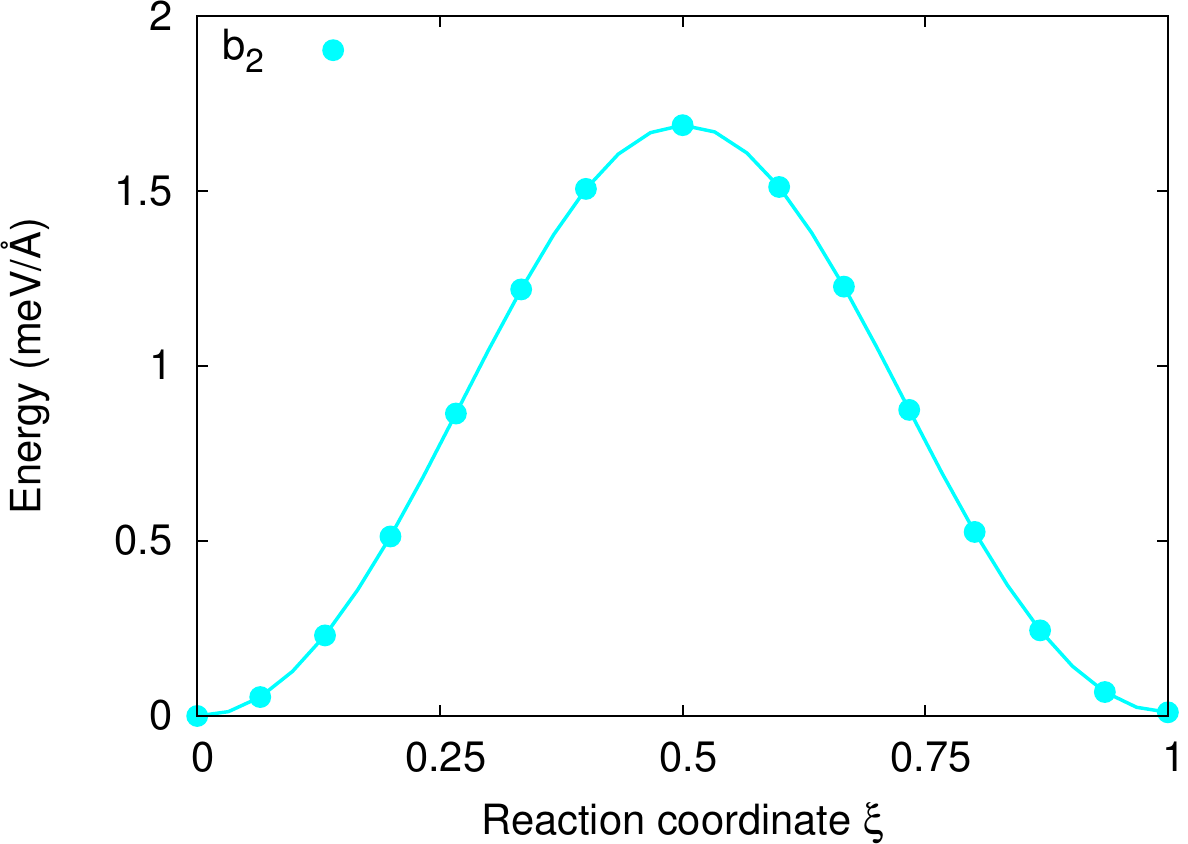}
		}
		\caption{Migration energies of the various disconnections under zero applied stress.
	}
	\label{fig:Migration_barriers}
\end{figure*}

Migration energies are calculated with the nudged elastic band (NEB) method.\cite{Henkelman2000}
The initial and final states of the NEB chain correspond to the same disconnection dipole with a length $\ell$ differing by one periodicity vector. 
The dipoles are chosen large enough so that this length variation leads to a negligible variation of the elastic energy along the path. 
The energy variation given by the NEB calculation then corresponds directly to the disconnection migration energy.
When dipoles are asymmetrical, we performed NEB calculations for both disconnections separately in order to obtain the two different migration barriers.
Calculations are performed under zero applied stress.
The NEB calculations were performed with the same periodicity vectors for each replica.
It implies, through equation \ref{eqn:boxes_strain}, that the applied stress along the path is not rigorously constant.
The variations are however small and we chose in practice the periodicity vectors canceling the stress for a disconnection dipole halfway between the initial and final states. 

Finally, one should point out that the length of the disconnection dipoles along their line direction are minimal and thus does not allow for the formation of kinks.\cite{Combe2016}
The obtained energy barriers therefore correspond to the 1D migration of the disconnections.
They constitute an upper limit of the 3D energy barriers when disconnection migration proceeds via the nucleation of double kinks
and are a necessary input for line tension or elastic models predicting kink pair formation.\cite{Proville2013,Kraych2016}

Results displayed in figure \ref{fig:Migration_barriers} show the barriers for the different disconnections on the four twin boundaries.
It can be observed that the energetic landscapes are different for all disconnection cores.
On the other hand, the migration energies are in general much lower than the formation energies (Fig. \ref{fig:Isolated_dipoles}).

For the \hkl{10-11} twin system, only the \burg{4} disconnection is compatible with the experimental twinning elements 
(twinning direction, intensity and activity under \hkl<c> compression). 
This disconnection has nevertheless a migration energy one order magnitude higher than the 
\burg{1}, \burg{2} and \burg{3m} disconnections, which can also exist on this twinning plane. 
The migration energy therefor does not appear as a key factor in the selection of the disconnection responsible for twin growth.
The same conclusion is reached for the \hkl{11-22} twin system where the \burg{3} disconnections,
which are the only ones that can account for twin growth under compression, do not have the lowest migration energy.

\subsection{Peierls stress}
\label{subsec:tau_P}

\begin{figure}[!b]
		\includegraphics[width=.8\linewidth]{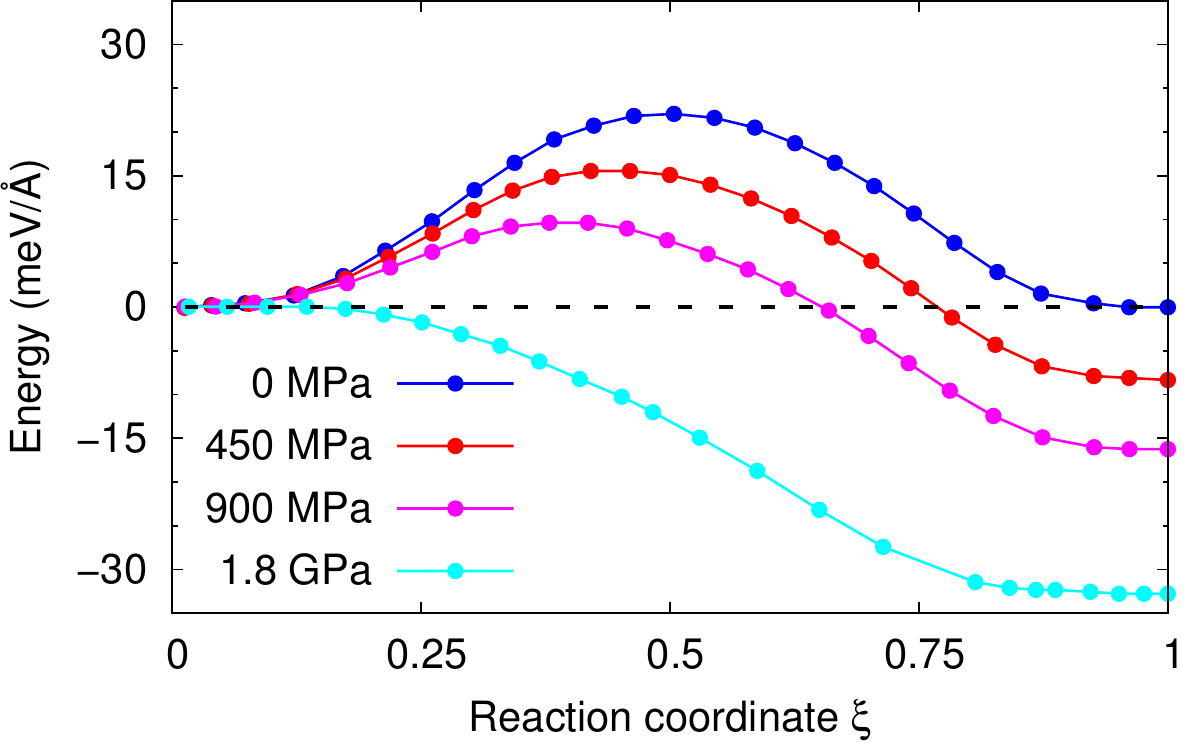}
	\caption{Migration barrier for the \burger{3}{} disconnection of the \hkl{11-22} twin system under various applied stresses.}
	\label{fig:NEB_stress_11_22}
\end{figure}

In addition to migration energies, the Peierls stress $\tau_{\rm P}$ has been determined by applying gradually a strain 
producing a resolved shear stress  on the disconnections in their glide direction (see Tab. \ref{tab:E_and_k}).
The Peierls stress corresponds to the critical resolved shear stress for which a disconnection of the dipole moves by at least one Peierls valley without the help of thermal fluctuations.
We checked by performing NEB calculations under applied stresses that the Peierls stress thus obtained corresponds to the applied stress which cancels the disconnection migration barrier. 
An example is shown in figure \ref{fig:NEB_stress_11_22} for the \burger{3}{S} disconnection of the \hkl{11-22} twin system. 
The direct calculation leads to a Peierls stress of 1.87 GPa (Tab. \ref{tab:E_and_k}), 
in agreement with the NEB calculation showing almost no energy barrier for an applied stress of 1.8 GPa (Fig. \ref{fig:NEB_stress_11_22}).
We note that for asymmetric dipoles, only the lowest Peierls stress is obtained.

As for the migration energy, the obtained Peierls stresses (Tab. \ref{tab:E_and_k}) exhibit a large variability amongst disconnections. 
For the twinning systems where several different disconnections exist, the ones corresponding to the experimental twinning elements,
\burg{4} on \hkl{10-11} and \burg{3} on \hkl{11-22}, are not the ones with the lowest Peierls stress.

\begin{figure}[!bt]
	\begin{center}
		\includegraphics[width=.8\linewidth]{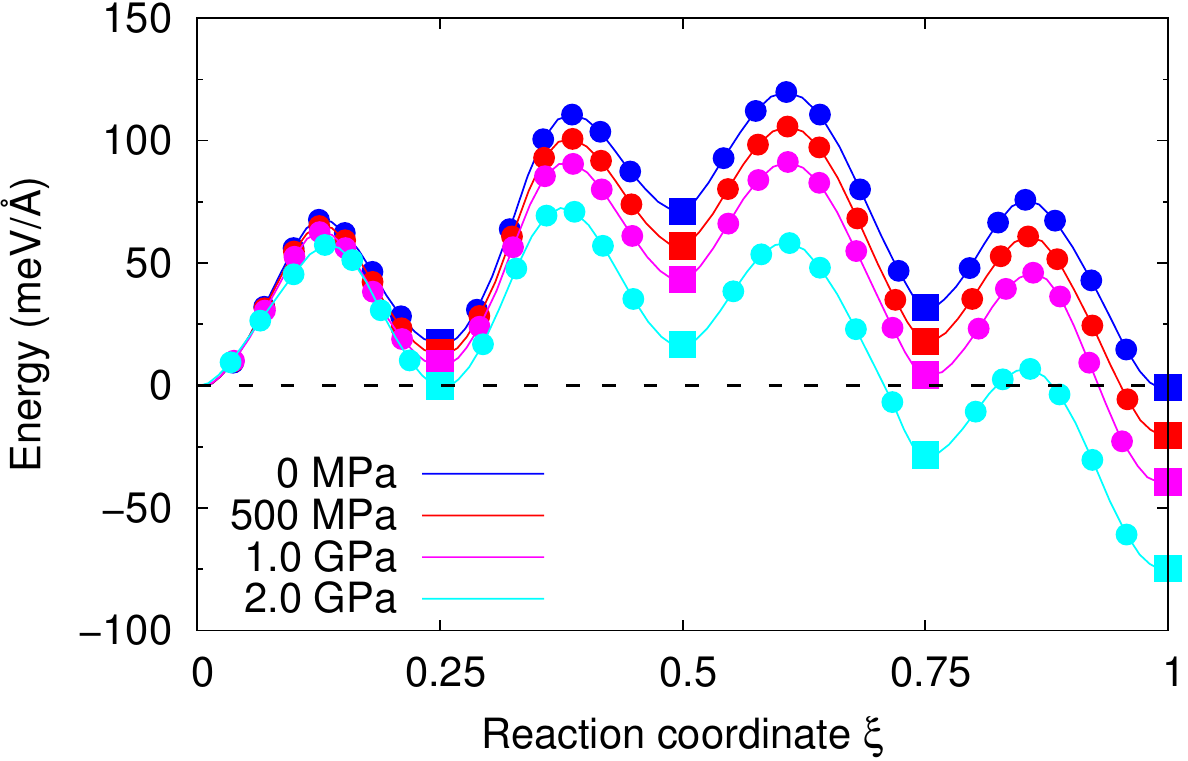}
	\end{center}
	\caption{Migration barrier for the \burger{4}{} disconnection of the \hkl{10-11} twin system under various applied stress. 
	The squares correspond to stable intermediate configurations relaxed independently.
	}
	\label{fig:NEB_stress_10_11}
\end{figure}

The \burg{4} disconnection, which is responsible for \hkl{10-11} twin growth, shows a particular behavior. 
It was not possible to make this disconnection glide, even for an applied stress as large as 2 GPa. 
This is confirmed by NEB calculations under applied stresses (Fig. \ref{fig:NEB_stress_10_11}),
showing that an energy barrier still remains with an applied stress of 2 GPa.
The motion of the \burg{4} disconnection along the \hkl{10-11} plane takes place in four successive steps,
with three intermediate metastable configurations. 
Despite a different sensitivity to the applied stress, none of the energy barriers associated with these successive steps 
disappears under applied stress. 
In particular, the first barrier appears uncoupled with the applied stress.
Migration of the \burg{4} disconnection will therefore require thermal activation which may be the reason why \hkl{10-11} twinning is active only at high temperature, \cite{Yoo1991} typically above $400\,^{\circ}\mathrm{C}$ in Ti.

\section{Conclusions}
\label{sec:Conclusions}

The two stages of disconnection nucleation relevant for twin thickening, i.e. the formation of disconnection dipoles and migration of disconnections along twin boundaries, have been modeled in hcp zirconium.
Using an EAM potential validated on \ai calculations, four different twin systems were studied at an atomic scale. 

A precise definition of the disconnection dipole energy is allowed thanks to an approach coupling elasticity and atomistic developed here.
It was shown that the formation energy of a disconnection dipole is composed of an elastic contribution, 
which depends on the surrounding microstructure, 
and a core energy, which is an intrinsic property of the disconnections. 
Computing separately the elastic contribution, thanks to an analogy with an Eshelby inclusion,
core energies could be extracted from atomistic simulations to define the formation energy of an isolated dipole.
This method was shown to reproduce well the behavior of large systems despite being parameterized on simulations of small sizes.
Of course, once the core energy is known, it becomes possible to compute formation energies
in more complex environments, like stress concentrations, 
or for more complex geometries, like disconnections loops.\cite{Luque2014}
In the latter case, the variation  of the core energy with the disconnection character
will need to be known.
The present study was restricted to disconnections normal to the twinning direction,
that usually have a pure edge character. 
But the same approach can be applied to various line orientations of the disconnection dipole, 
to obtain the variation of the core energy with the disconnection character.

Considering the different disconnections that can exist on twin boundaries, 
it has been shown that the most stable is not always the one with the smallest Burgers vector, 
i.e. the one with the lowest elastic energy.  Because of the core contribution, 
it sometimes happens that a disconnection with a larger Burgers vector 
is more stable over a wide range of dipole lengths.
It appears that the energy scale controlling disconnection formation and migration in Zr is much higher for the \hkl{10-11} and \hkl{11-22} twin systems active under compression than for the \hkl{10-12} and \hkl{11-21} tensile twin systems.
Considering both the core and elastic contributions, atomistic simulations predict that the twinning modes experimentally observed in hcp Zr correspond to the disconnections with the lowest formation energy. 
On the other hand, these disconnections are not necessarily the ones migrating the most easily. 
They generally neither have the lowest migration energy nor the lowest Peierls stress. 
As a consequence the nucleation of disconnection dipoles appears more critical than their  migration for twin mode selection in Zr. 


\begin{acknowledgments}
  This work was performed using HPC resources from GENCI-CINES and GENCI-CCRT (Grants 2016-096847). 
  DR acknowledges support from LABEX iMUST (ANR-10-LABX-0064) of Université de Lyon (program ``Investissements d'Avenir'', ANR-11-IDEX-0007)
\end{acknowledgments}

\bibliographystyle{apsrev4-1}
\bibliography{mackain2017}

\begin{thebibliography}{63}%
\makeatletter
\providecommand \@ifxundefined [1]{%
 \@ifx{#1\undefined}
}%
\providecommand \@ifnum [1]{%
 \ifnum #1\expandafter \@firstoftwo
 \else \expandafter \@secondoftwo
 \fi
}%
\providecommand \@ifx [1]{%
 \ifx #1\expandafter \@firstoftwo
 \else \expandafter \@secondoftwo
 \fi
}%
\providecommand \natexlab [1]{#1}%
\providecommand \enquote  [1]{``#1''}%
\providecommand \bibnamefont  [1]{#1}%
\providecommand \bibfnamefont [1]{#1}%
\providecommand \citenamefont [1]{#1}%
\providecommand \href@noop [0]{\@secondoftwo}%
\providecommand \href [0]{\begingroup \@sanitize@url \@href}%
\providecommand \@href[1]{\@@startlink{#1}\@@href}%
\providecommand \@@href[1]{\endgroup#1\@@endlink}%
\providecommand \@sanitize@url [0]{\catcode `\\12\catcode `\$12\catcode
  `\&12\catcode `\#12\catcode `\^12\catcode `\_12\catcode `\%12\relax}%
\providecommand \@@startlink[1]{}%
\providecommand \@@endlink[0]{}%
\providecommand \url  [0]{\begingroup\@sanitize@url \@url }%
\providecommand \@url [1]{\endgroup\@href {#1}{\urlprefix }}%
\providecommand \urlprefix  [0]{URL }%
\providecommand \Eprint [0]{\href }%
\providecommand \doibase [0]{http://dx.doi.org/}%
\providecommand \selectlanguage [0]{\@gobble}%
\providecommand \bibinfo  [0]{\@secondoftwo}%
\providecommand \bibfield  [0]{\@secondoftwo}%
\providecommand \translation [1]{[#1]}%
\providecommand \BibitemOpen [0]{}%
\providecommand \bibitemStop [0]{}%
\providecommand \bibitemNoStop [0]{.\EOS\space}%
\providecommand \EOS [0]{\spacefactor3000\relax}%
\providecommand \BibitemShut  [1]{\csname bibitem#1\endcsname}%
\let\auto@bib@innerbib\@empty
\bibitem [{\citenamefont {Partridge}(1967)}]{Partridge1967}%
  \BibitemOpen
  \bibfield  {author} {\bibinfo {author} {\bibfnamefont {P.~G.}\ \bibnamefont
  {Partridge}},\ }\href@noop {} {\bibfield  {journal} {\bibinfo  {journal}
  {Metall. Rev.}\ }\textbf {\bibinfo {volume} {12 (1)}},\ \bibinfo {pages} {169
  } (\bibinfo {year} {1967})}\BibitemShut {NoStop}%
\bibitem [{\citenamefont {Christian}\ and\ \citenamefont
  {Mahajan}(1995)}]{Christian1995}%
  \BibitemOpen
  \bibfield  {author} {\bibinfo {author} {\bibfnamefont {J.~W.}\ \bibnamefont
  {Christian}}\ and\ \bibinfo {author} {\bibfnamefont {S.}~\bibnamefont
  {Mahajan}},\ }\href {\doibase 10.1016/0079-6425(94)00007-7} {\bibfield
  {journal} {\bibinfo  {journal} {Prog. Mater. Sci.}\ }\textbf {\bibinfo
  {volume} {39}},\ \bibinfo {pages} {1} (\bibinfo {year} {1995})}\BibitemShut
  {NoStop}%
\bibitem [{\citenamefont {Hirth}\ and\ \citenamefont
  {Lothe}(1982)}]{Hirth1982}%
  \BibitemOpen
  \bibfield  {author} {\bibinfo {author} {\bibfnamefont {J.~P.}\ \bibnamefont
  {Hirth}}\ and\ \bibinfo {author} {\bibfnamefont {J.}~\bibnamefont {Lothe}},\
  }\href@noop {} {\emph {\bibinfo {title} {Theory of disclocations}}}\
  (\bibinfo  {publisher} {John Wiley \& Sons},\ \bibinfo {year}
  {1982})\BibitemShut {NoStop}%
\bibitem [{\citenamefont {Yoo}(1981)}]{Yoo1981}%
  \BibitemOpen
  \bibfield  {author} {\bibinfo {author} {\bibfnamefont {M.~H.}\ \bibnamefont
  {Yoo}},\ }\href {\doibase 10.1007/BF02648537} {\bibfield  {journal} {\bibinfo
   {journal} {Metall. Trans. A}\ }\textbf {\bibinfo {volume} {12}},\ \bibinfo
  {pages} {409} (\bibinfo {year} {1981})}\BibitemShut {NoStop}%
\bibitem [{\citenamefont {Beyerlein}\ \emph {et~al.}(2014)\citenamefont
  {Beyerlein}, \citenamefont {Zhang},\ and\ \citenamefont
  {Misra}}]{Beyerlein2014}%
  \BibitemOpen
  \bibfield  {author} {\bibinfo {author} {\bibfnamefont {I.~J.}\ \bibnamefont
  {Beyerlein}}, \bibinfo {author} {\bibfnamefont {X.}~\bibnamefont {Zhang}}, \
  and\ \bibinfo {author} {\bibfnamefont {A.}~\bibnamefont {Misra}},\ }\href
  {\doibase 10.1146/annurev-matsci-070813-113304} {\bibfield  {journal}
  {\bibinfo  {journal} {Annual Review of Materials Research}\ }\textbf
  {\bibinfo {volume} {44}},\ \bibinfo {pages} {329 } (\bibinfo {year}
  {2014})}\BibitemShut {NoStop}%
\bibitem [{\citenamefont {Hirth}\ \emph {et~al.}(2016)\citenamefont {Hirth},
  \citenamefont {Wang},\ and\ \citenamefont {Tomé}}]{Hirth2016}%
  \BibitemOpen
  \bibfield  {author} {\bibinfo {author} {\bibfnamefont {J.~P.}\ \bibnamefont
  {Hirth}}, \bibinfo {author} {\bibfnamefont {J.}~\bibnamefont {Wang}}, \ and\
  \bibinfo {author} {\bibfnamefont {C.~N.}\ \bibnamefont {Tomé}},\ }\href
  {\doibase 10.1016/j.pmatsci.2016.07.003} {\bibfield  {journal} {\bibinfo
  {journal} {Prog. Mater. Sci.}\ }\textbf {\bibinfo {volume} {83}},\ \bibinfo
  {pages} {417} (\bibinfo {year} {2016})}\BibitemShut {NoStop}%
\bibitem [{\citenamefont {Serra}\ and\ \citenamefont
  {Bacon}(1995)}]{Serra1995}%
  \BibitemOpen
  \bibfield  {author} {\bibinfo {author} {\bibfnamefont {A.}~\bibnamefont
  {Serra}}\ and\ \bibinfo {author} {\bibfnamefont {D.}~\bibnamefont {Bacon}},\
  }\href {\doibase 10.1016/0956-7151(95)00128-I} {\bibfield  {journal}
  {\bibinfo  {journal} {Acta Metall. Mater.}\ }\textbf {\bibinfo {volume}
  {43}},\ \bibinfo {pages} {4465} (\bibinfo {year} {1995})}\BibitemShut
  {NoStop}%
\bibitem [{\citenamefont {Serra}\ and\ \citenamefont
  {Bacon}(1996)}]{Serra1996}%
  \BibitemOpen
  \bibfield  {author} {\bibinfo {author} {\bibfnamefont {A.}~\bibnamefont
  {Serra}}\ and\ \bibinfo {author} {\bibfnamefont {D.~J.}\ \bibnamefont
  {Bacon}},\ }\href {\doibase 10.1080/01418619608244386} {\bibfield  {journal}
  {\bibinfo  {journal} {Philos. Mag. A}\ }\textbf {\bibinfo {volume} {73}},\
  \bibinfo {pages} {333} (\bibinfo {year} {1996})}\BibitemShut {NoStop}%
\bibitem [{\citenamefont {Wang}\ \emph {et~al.}(2012)\citenamefont {Wang},
  \citenamefont {Beyerlein},\ and\ \citenamefont {Hirth}}]{Wang2012}%
  \BibitemOpen
  \bibfield  {author} {\bibinfo {author} {\bibfnamefont {J.}~\bibnamefont
  {Wang}}, \bibinfo {author} {\bibfnamefont {I.~J.}\ \bibnamefont {Beyerlein}},
  \ and\ \bibinfo {author} {\bibfnamefont {J.~P.}\ \bibnamefont {Hirth}},\
  }\href {\doibase 10.1088/0965-0393/20/2/024001} {\bibfield  {journal}
  {\bibinfo  {journal} {Modelling Simul. Mater. Sci. Eng.}\ }\textbf {\bibinfo
  {volume} {20}},\ \bibinfo {pages} {024001} (\bibinfo {year}
  {2012})}\BibitemShut {NoStop}%
\bibitem [{\citenamefont {Kadiri}\ \emph {et~al.}(2015)\citenamefont {Kadiri},
  \citenamefont {Barrett}, \citenamefont {Wang},\ and\ \citenamefont
  {Tom{\'e}}}]{Kadiri2015}%
  \BibitemOpen
  \bibfield  {author} {\bibinfo {author} {\bibfnamefont {H.~E.}\ \bibnamefont
  {Kadiri}}, \bibinfo {author} {\bibfnamefont {C.~D.}\ \bibnamefont {Barrett}},
  \bibinfo {author} {\bibfnamefont {J.}~\bibnamefont {Wang}}, \ and\ \bibinfo
  {author} {\bibfnamefont {C.~N.}\ \bibnamefont {Tom{\'e}}},\ }\href {\doibase
  http://dx.doi.org/10.1016/j.actamat.2014.11.033} {\bibfield  {journal}
  {\bibinfo  {journal} {Acta Mater.}\ }\textbf {\bibinfo {volume} {85}},\
  \bibinfo {pages} {354 } (\bibinfo {year} {2015})}\BibitemShut {NoStop}%
\bibitem [{\citenamefont {Fan}\ \emph {et~al.}(2015)\citenamefont {Fan},
  \citenamefont {Aubry}, \citenamefont {Arsenlis},\ and\ \citenamefont
  {El-Awady}}]{Fan2015}%
  \BibitemOpen
  \bibfield  {author} {\bibinfo {author} {\bibfnamefont {H.}~\bibnamefont
  {Fan}}, \bibinfo {author} {\bibfnamefont {S.}~\bibnamefont {Aubry}}, \bibinfo
  {author} {\bibfnamefont {A.}~\bibnamefont {Arsenlis}}, \ and\ \bibinfo
  {author} {\bibfnamefont {J.}~\bibnamefont {El-Awady}},\ }\href {\doibase
  http://dx.doi.org/10.1016/j.actamat.2015.03.039} {\bibfield  {journal}
  {\bibinfo  {journal} {Acta Mater.}\ }\textbf {\bibinfo {volume} {92}},\
  \bibinfo {pages} {126 } (\bibinfo {year} {2015})}\BibitemShut {NoStop}%
\bibitem [{\citenamefont {Fan}\ \emph {et~al.}(2016)\citenamefont {Fan},
  \citenamefont {Aubry}, \citenamefont {Arsenlis},\ and\ \citenamefont
  {El-Awady}}]{Fan2016}%
  \BibitemOpen
  \bibfield  {author} {\bibinfo {author} {\bibfnamefont {H.}~\bibnamefont
  {Fan}}, \bibinfo {author} {\bibfnamefont {S.}~\bibnamefont {Aubry}}, \bibinfo
  {author} {\bibfnamefont {A.}~\bibnamefont {Arsenlis}}, \ and\ \bibinfo
  {author} {\bibfnamefont {J.}~\bibnamefont {El-Awady}},\ }\href {\doibase
  doi:10.1016/j.scriptamat.2015.09.008} {\bibfield  {journal} {\bibinfo
  {journal} {Scripta Mater.}\ }\textbf {\bibinfo {volume} {112}},\ \bibinfo
  {pages} {50 } (\bibinfo {year} {2016})}\BibitemShut {NoStop}%
\bibitem [{\citenamefont {Wang}\ and\ \citenamefont {Agnew}(2016)}]{Wang2016}%
  \BibitemOpen
  \bibfield  {author} {\bibinfo {author} {\bibfnamefont {F.}~\bibnamefont
  {Wang}}\ and\ \bibinfo {author} {\bibfnamefont {S.~R.}\ \bibnamefont
  {Agnew}},\ }\href {\doibase http://dx.doi.org/10.1016/j.ijplas.2016.01.012}
  {\bibfield  {journal} {\bibinfo  {journal} {Int. J. Plast.}\ }\textbf
  {\bibinfo {volume} {81}},\ \bibinfo {pages} {63 } (\bibinfo {year}
  {2016})}\BibitemShut {NoStop}%
\bibitem [{\citenamefont {Ghazisaeidi}\ \emph {et~al.}(2014)\citenamefont
  {Ghazisaeidi}, \citenamefont {Hector},\ and\ \citenamefont
  {Curtin}}]{Ghazisaeidi2014}%
  \BibitemOpen
  \bibfield  {author} {\bibinfo {author} {\bibfnamefont {M.}~\bibnamefont
  {Ghazisaeidi}}, \bibinfo {author} {\bibfnamefont {L.~G.}\ \bibnamefont
  {Hector}}, \ and\ \bibinfo {author} {\bibfnamefont {W.~A.}\ \bibnamefont
  {Curtin}},\ }\href {\doibase 10.1016/j.actamat.2014.07.045} {\bibfield
  {journal} {\bibinfo  {journal} {Acta Mater.}\ }\textbf {\bibinfo {volume}
  {80}},\ \bibinfo {pages} {278 } (\bibinfo {year} {2014})}\BibitemShut
  {NoStop}%
\bibitem [{\citenamefont {Luque}\ \emph {et~al.}(2014)\citenamefont {Luque},
  \citenamefont {Ghazisaeidi},\ and\ \citenamefont {Curtin}}]{Luque2014}%
  \BibitemOpen
  \bibfield  {author} {\bibinfo {author} {\bibfnamefont {A.}~\bibnamefont
  {Luque}}, \bibinfo {author} {\bibfnamefont {M.}~\bibnamefont {Ghazisaeidi}},
  \ and\ \bibinfo {author} {\bibfnamefont {W.~A.}\ \bibnamefont {Curtin}},\
  }\href {\doibase http://dx.doi.org/10.1016/j.actamat.2014.08.052} {\bibfield
  {journal} {\bibinfo  {journal} {Acta Mater.}\ }\textbf {\bibinfo {volume}
  {81}},\ \bibinfo {pages} {442 } (\bibinfo {year} {2014})}\BibitemShut
  {NoStop}%
\bibitem [{\citenamefont {Serra}\ \emph {et~al.}(1988)\citenamefont {Serra},
  \citenamefont {Bacon},\ and\ \citenamefont {Pond}}]{Serra1988}%
  \BibitemOpen
  \bibfield  {author} {\bibinfo {author} {\bibfnamefont {A.}~\bibnamefont
  {Serra}}, \bibinfo {author} {\bibfnamefont {D.~J.}\ \bibnamefont {Bacon}}, \
  and\ \bibinfo {author} {\bibfnamefont {R.~C.}\ \bibnamefont {Pond}},\ }\href
  {\doibase 10.1016/0001-6160(88)90054-5} {\bibfield  {journal} {\bibinfo
  {journal} {Acta Metall.}\ }\textbf {\bibinfo {volume} {36}},\ \bibinfo
  {pages} {3183} (\bibinfo {year} {1988})}\BibitemShut {NoStop}%
\bibitem [{\citenamefont {Serra}\ \emph {et~al.}(1991)\citenamefont {Serra},
  \citenamefont {Pond},\ and\ \citenamefont {Bacon}}]{Serra1991}%
  \BibitemOpen
  \bibfield  {author} {\bibinfo {author} {\bibfnamefont {A.}~\bibnamefont
  {Serra}}, \bibinfo {author} {\bibfnamefont {R.~C.}\ \bibnamefont {Pond}}, \
  and\ \bibinfo {author} {\bibfnamefont {D.~J.}\ \bibnamefont {Bacon}},\ }\href
  {\doibase 10.1016/0956-7151(91)90232-P} {\bibfield  {journal} {\bibinfo
  {journal} {Acta Metall. Mater.}\ }\textbf {\bibinfo {volume} {39}},\ \bibinfo
  {pages} {1469} (\bibinfo {year} {1991})}\BibitemShut {NoStop}%
\bibitem [{\citenamefont {Bacon}\ and\ \citenamefont
  {Vitek}(2002)}]{Bacon2002}%
  \BibitemOpen
  \bibfield  {author} {\bibinfo {author} {\bibfnamefont {D.~J.}\ \bibnamefont
  {Bacon}}\ and\ \bibinfo {author} {\bibfnamefont {V.}~\bibnamefont {Vitek}},\
  }\href {\doibase 10.1007/s11661-002-0138-x} {\bibfield  {journal} {\bibinfo
  {journal} {Metall. Mater. Trans. A}\ }\textbf {\bibinfo {volume} {33}},\
  \bibinfo {pages} {721} (\bibinfo {year} {2002})}\BibitemShut {NoStop}%
\bibitem [{\citenamefont {Lemaignan}(2012)}]{Lemaignan2012}%
  \BibitemOpen
  \bibfield  {author} {\bibinfo {author} {\bibfnamefont {C.}~\bibnamefont
  {Lemaignan}},\ }\enquote {\bibinfo {title} {Zirconium alloys: Properties and
  characteristics},}\ in\ \href {\doibase 10.1016/B978-0-08-056033-5.00015-X}
  {\emph {\bibinfo {booktitle} {Comprehensive Nuclear Materials}}},\ \bibinfo
  {editor} {edited by\ \bibinfo {editor} {\bibfnamefont {R.~J.~M.}\
  \bibnamefont {Konings}}, \bibinfo {editor} {\bibfnamefont {T.~R.}\
  \bibnamefont {Allen}}, \bibinfo {editor} {\bibfnamefont {R.~E.}\ \bibnamefont
  {Stoller}}, \ and\ \bibinfo {editor} {\bibfnamefont {S.}~\bibnamefont
  {Yamanaka}}}\ (\bibinfo  {publisher} {Elsevier},\ \bibinfo {year} {2012})\
  Chap.\ \bibinfo {chapter} {2.07}, pp.\ \bibinfo {pages} {217 --
  232}\BibitemShut {NoStop}%
\bibitem [{\citenamefont {Yoo}\ and\ \citenamefont {Lee}(1991)}]{Yoo1991}%
  \BibitemOpen
  \bibfield  {author} {\bibinfo {author} {\bibfnamefont {M.~H.}\ \bibnamefont
  {Yoo}}\ and\ \bibinfo {author} {\bibfnamefont {J.~K.}\ \bibnamefont {Lee}},\
  }\href {\doibase 10.1080/01418619108213931} {\bibfield  {journal} {\bibinfo
  {journal} {Philos. Mag. A}\ }\textbf {\bibinfo {volume} {63}},\ \bibinfo
  {pages} {987} (\bibinfo {year} {1991})}\BibitemShut {NoStop}%
\bibitem [{\citenamefont {Mendelev}\ and\ \citenamefont
  {Ackland}(2007)}]{Mendelev2007}%
  \BibitemOpen
  \bibfield  {author} {\bibinfo {author} {\bibfnamefont {M.~I.}\ \bibnamefont
  {Mendelev}}\ and\ \bibinfo {author} {\bibfnamefont {G.~J.}\ \bibnamefont
  {Ackland}},\ }\href {\doibase 10.1080/09500830701191393} {\bibfield
  {journal} {\bibinfo  {journal} {Philos. Mag. Lett.}\ }\textbf {\bibinfo
  {volume} {87}},\ \bibinfo {pages} {349 } (\bibinfo {year}
  {2007})}\BibitemShut {NoStop}%
\bibitem [{\citenamefont {Khater}\ and\ \citenamefont
  {Bacon}(2010)}]{Khater2010}%
  \BibitemOpen
  \bibfield  {author} {\bibinfo {author} {\bibfnamefont {H.~A.}\ \bibnamefont
  {Khater}}\ and\ \bibinfo {author} {\bibfnamefont {D.~J.}\ \bibnamefont
  {Bacon}},\ }\href {\doibase 10.1016/j.actamat.2010.01.028} {\bibfield
  {journal} {\bibinfo  {journal} {Acta Mater.}\ }\textbf {\bibinfo {volume}
  {58}},\ \bibinfo {pages} {2978} (\bibinfo {year} {2010})}\BibitemShut
  {NoStop}%
\bibitem [{\citenamefont {Clouet}(2012)}]{Clouet2012}%
  \BibitemOpen
  \bibfield  {author} {\bibinfo {author} {\bibfnamefont {E.}~\bibnamefont
  {Clouet}},\ }\href {\doibase 10.1103/PhysRevB.86.144104} {\bibfield
  {journal} {\bibinfo  {journal} {Phys. Rev. B}\ }\textbf {\bibinfo {volume}
  {86}},\ \bibinfo {pages} {144104} (\bibinfo {year} {2012})}\BibitemShut
  {NoStop}%
\bibitem [{\citenamefont {Chaari}\ \emph
  {et~al.}(2014{\natexlab{a}})\citenamefont {Chaari}, \citenamefont {Clouet},\
  and\ \citenamefont {Rodney}}]{Chaari2014}%
  \BibitemOpen
  \bibfield  {author} {\bibinfo {author} {\bibfnamefont {N.}~\bibnamefont
  {Chaari}}, \bibinfo {author} {\bibfnamefont {E.}~\bibnamefont {Clouet}}, \
  and\ \bibinfo {author} {\bibfnamefont {D.}~\bibnamefont {Rodney}},\ }\href
  {\doibase 10.1103/PhysRevLett.112.075504} {\bibfield  {journal} {\bibinfo
  {journal} {Phys. Rev. Lett.}\ }\textbf {\bibinfo {volume} {112}},\ \bibinfo
  {pages} {075504} (\bibinfo {year} {2014}{\natexlab{a}})}\BibitemShut
  {NoStop}%
\bibitem [{\citenamefont {Chaari}\ \emph
  {et~al.}(2014{\natexlab{b}})\citenamefont {Chaari}, \citenamefont {Clouet},\
  and\ \citenamefont {Rodney}}]{Chaari2014a}%
  \BibitemOpen
  \bibfield  {author} {\bibinfo {author} {\bibfnamefont {N.}~\bibnamefont
  {Chaari}}, \bibinfo {author} {\bibfnamefont {E.}~\bibnamefont {Clouet}}, \
  and\ \bibinfo {author} {\bibfnamefont {D.}~\bibnamefont {Rodney}},\ }\href
  {\doibase 10.1007/s11661-014-2568-7} {\bibfield  {journal} {\bibinfo
  {journal} {Metall. Mater. Trans. A}\ }\textbf {\bibinfo {volume} {45}},\
  \bibinfo {pages} {5898} (\bibinfo {year} {2014}{\natexlab{b}})}\BibitemShut
  {NoStop}%
\bibitem [{\citenamefont {Lu}\ \emph {et~al.}(2015{\natexlab{a}})\citenamefont
  {Lu}, \citenamefont {Chernatynskiy}, \citenamefont {Noordhoek}, \citenamefont
  {Sinnott},\ and\ \citenamefont {Phillpot}}]{Lu2015a}%
  \BibitemOpen
  \bibfield  {author} {\bibinfo {author} {\bibfnamefont {Z.}~\bibnamefont
  {Lu}}, \bibinfo {author} {\bibfnamefont {A.}~\bibnamefont {Chernatynskiy}},
  \bibinfo {author} {\bibfnamefont {M.~J.}\ \bibnamefont {Noordhoek}}, \bibinfo
  {author} {\bibfnamefont {S.~B.}\ \bibnamefont {Sinnott}}, \ and\ \bibinfo
  {author} {\bibfnamefont {S.~R.}\ \bibnamefont {Phillpot}},\ }\href {\doibase
  10.1016/j.jnucmat.2015.10.043} {\bibfield  {journal} {\bibinfo  {journal} {J.
  Nucl. Mater.}\ }\textbf {\bibinfo {volume} {467}},\ \bibinfo {pages} {742 }
  (\bibinfo {year} {2015}{\natexlab{a}})}\BibitemShut {NoStop}%
\bibitem [{\citenamefont {Lu}\ \emph {et~al.}(2015{\natexlab{b}})\citenamefont
  {Lu}, \citenamefont {Noordhoek}, \citenamefont {Chernatynskiy}, \citenamefont
  {Sinnott},\ and\ \citenamefont {Phillpot}}]{Lu2015}%
  \BibitemOpen
  \bibfield  {author} {\bibinfo {author} {\bibfnamefont {Z.}~\bibnamefont
  {Lu}}, \bibinfo {author} {\bibfnamefont {M.~J.}\ \bibnamefont {Noordhoek}},
  \bibinfo {author} {\bibfnamefont {A.}~\bibnamefont {Chernatynskiy}}, \bibinfo
  {author} {\bibfnamefont {S.~B.}\ \bibnamefont {Sinnott}}, \ and\ \bibinfo
  {author} {\bibfnamefont {S.~R.}\ \bibnamefont {Phillpot}},\ }\href {\doibase
  10.1016/j.jnucmat.2015.03.048} {\bibfield  {journal} {\bibinfo  {journal} {J.
  Nucl. Mater.}\ }\textbf {\bibinfo {volume} {462}},\ \bibinfo {pages} {147 }
  (\bibinfo {year} {2015}{\natexlab{b}})}\BibitemShut {NoStop}%
\bibitem [{\citenamefont {Szewc}\ \emph {et~al.}(2016)\citenamefont {Szewc},
  \citenamefont {Pizzagalli}, \citenamefont {Brochard},\ and\ \citenamefont
  {Clouet}}]{Szewc2016}%
  \BibitemOpen
  \bibfield  {author} {\bibinfo {author} {\bibfnamefont {W.}~\bibnamefont
  {Szewc}}, \bibinfo {author} {\bibfnamefont {L.}~\bibnamefont {Pizzagalli}},
  \bibinfo {author} {\bibfnamefont {S.}~\bibnamefont {Brochard}}, \ and\
  \bibinfo {author} {\bibfnamefont {E.}~\bibnamefont {Clouet}},\ }\href
  {\doibase http://dx.doi.org/10.1016/j.actamat.2016.05.025} {\bibfield
  {journal} {\bibinfo  {journal} {Acta Mater.}\ }\textbf {\bibinfo {volume}
  {114}},\ \bibinfo {pages} {126 } (\bibinfo {year} {2016})}\BibitemShut
  {NoStop}%
\bibitem [{\citenamefont {Kresse}\ and\ \citenamefont
  {Furthmüller}(1996)}]{Kresse1996}%
  \BibitemOpen
  \bibfield  {author} {\bibinfo {author} {\bibfnamefont {G.}~\bibnamefont
  {Kresse}}\ and\ \bibinfo {author} {\bibfnamefont {J.}~\bibnamefont
  {Furthmüller}},\ }\href {\doibase
  http://dx.doi.org/10.1016/0927-0256(96)00008-0} {\bibfield  {journal}
  {\bibinfo  {journal} {Comp. Mat. Sci.}\ }\textbf {\bibinfo {volume} {6}},\
  \bibinfo {pages} {15 } (\bibinfo {year} {1996})}\BibitemShut {NoStop}%
\bibitem [{\citenamefont {Perdew}\ \emph {et~al.}(1996)\citenamefont {Perdew},
  \citenamefont {Burke},\ and\ \citenamefont {Ernzerhof}}]{Perdew1996}%
  \BibitemOpen
  \bibfield  {author} {\bibinfo {author} {\bibfnamefont {J.~P.}\ \bibnamefont
  {Perdew}}, \bibinfo {author} {\bibfnamefont {K.}~\bibnamefont {Burke}}, \
  and\ \bibinfo {author} {\bibfnamefont {M.}~\bibnamefont {Ernzerhof}},\ }\href
  {\doibase 10.1103/PhysRevLett.77.3865} {\bibfield  {journal} {\bibinfo
  {journal} {Phys. Rev. Lett.}\ }\textbf {\bibinfo {volume} {77}},\ \bibinfo
  {pages} {3865 } (\bibinfo {year} {1996})}\BibitemShut {NoStop}%
\bibitem [{\citenamefont {Kumar}\ \emph {et~al.}(2015)\citenamefont {Kumar},
  \citenamefont {Wang},\ and\ \citenamefont {Tomé}}]{Kumar2015}%
  \BibitemOpen
  \bibfield  {author} {\bibinfo {author} {\bibfnamefont {A.}~\bibnamefont
  {Kumar}}, \bibinfo {author} {\bibfnamefont {J.}~\bibnamefont {Wang}}, \ and\
  \bibinfo {author} {\bibfnamefont {C.~N.}\ \bibnamefont {Tomé}},\ }\href
  {\doibase http;//dx.doi.org/10.1016/j.actamat.2014.11.015} {\bibfield
  {journal} {\bibinfo  {journal} {Acta Mater.}\ }\textbf {\bibinfo {volume}
  {85}},\ \bibinfo {pages} {144 } (\bibinfo {year} {2015})}\BibitemShut
  {NoStop}%
\bibitem [{\citenamefont {Frank}(1965)}]{Frank1965}%
  \BibitemOpen
  \bibfield  {author} {\bibinfo {author} {\bibfnamefont {F.~C.}\ \bibnamefont
  {Frank}},\ }\href {\doibase 10.1107/s0365110x65002116} {\bibfield  {journal}
  {\bibinfo  {journal} {Acta Crystallogr.}\ }\textbf {\bibinfo {volume} {18}},\
  \bibinfo {pages} {862} (\bibinfo {year} {1965})}\BibitemShut {NoStop}%
\bibitem [{\citenamefont {Morris}\ \emph {et~al.}(1994)\citenamefont {Morris},
  \citenamefont {Ye}, \citenamefont {Ho}, \citenamefont {Chan},\ and\
  \citenamefont {Yoo}}]{Morris1994}%
  \BibitemOpen
  \bibfield  {author} {\bibinfo {author} {\bibfnamefont {J.~R.}\ \bibnamefont
  {Morris}}, \bibinfo {author} {\bibfnamefont {Y.~Y.}\ \bibnamefont {Ye}},
  \bibinfo {author} {\bibfnamefont {K.~M.}\ \bibnamefont {Ho}}, \bibinfo
  {author} {\bibfnamefont {C.~T.}\ \bibnamefont {Chan}}, \ and\ \bibinfo
  {author} {\bibfnamefont {M.~H.}\ \bibnamefont {Yoo}},\ }\href {\doibase
  10.1080/09500839408241591} {\bibfield  {journal} {\bibinfo  {journal}
  {Philos. Mag. Lett.}\ }\textbf {\bibinfo {volume} {69}},\ \bibinfo {pages}
  {189} (\bibinfo {year} {1994})}\BibitemShut {NoStop}%
\bibitem [{\citenamefont {Morris}\ \emph {et~al.}(1995)\citenamefont {Morris},
  \citenamefont {Ye}, \citenamefont {Ho}, \citenamefont {Chan},\ and\
  \citenamefont {Yoo}}]{Morris1995}%
  \BibitemOpen
  \bibfield  {author} {\bibinfo {author} {\bibfnamefont {J.~R.}\ \bibnamefont
  {Morris}}, \bibinfo {author} {\bibfnamefont {Y.~Y.}\ \bibnamefont {Ye}},
  \bibinfo {author} {\bibfnamefont {K.~M.}\ \bibnamefont {Ho}}, \bibinfo
  {author} {\bibfnamefont {C.~T.}\ \bibnamefont {Chan}}, \ and\ \bibinfo
  {author} {\bibfnamefont {M.~H.}\ \bibnamefont {Yoo}},\ }\href {\doibase
  10.1080/01418619508243798} {\bibfield  {journal} {\bibinfo  {journal}
  {Philos. Mag. A}\ }\textbf {\bibinfo {volume} {72}},\ \bibinfo {pages} {751}
  (\bibinfo {year} {1995})}\BibitemShut {NoStop}%
\bibitem [{\citenamefont {de~Jong}\ \emph {et~al.}(2015)\citenamefont
  {de~Jong}, \citenamefont {Kacher}, \citenamefont {Sluiter}, \citenamefont
  {Qi}, \citenamefont {Olmsted}, \citenamefont {van~de Walle}, \citenamefont
  {Morris}, \citenamefont {Minor},\ and\ \citenamefont {Asta}}]{Jong2015}%
  \BibitemOpen
  \bibfield  {author} {\bibinfo {author} {\bibfnamefont {M.}~\bibnamefont
  {de~Jong}}, \bibinfo {author} {\bibfnamefont {J.}~\bibnamefont {Kacher}},
  \bibinfo {author} {\bibfnamefont {M.~H.~F.}\ \bibnamefont {Sluiter}},
  \bibinfo {author} {\bibfnamefont {L.}~\bibnamefont {Qi}}, \bibinfo {author}
  {\bibfnamefont {D.~L.}\ \bibnamefont {Olmsted}}, \bibinfo {author}
  {\bibfnamefont {A.}~\bibnamefont {van~de Walle}}, \bibinfo {author}
  {\bibfnamefont {J.~W.~J.}\ \bibnamefont {Morris}}, \bibinfo {author}
  {\bibfnamefont {A.~M.}\ \bibnamefont {Minor}}, \ and\ \bibinfo {author}
  {\bibfnamefont {M.}~\bibnamefont {Asta}},\ }\href {\doibase
  10.1103/PhysRevLett.115.065501} {\bibfield  {journal} {\bibinfo  {journal}
  {Phys. Rev. Lett.}\ }\textbf {\bibinfo {volume} {115}},\ \bibinfo {pages}
  {065501} (\bibinfo {year} {2015})}\BibitemShut {NoStop}%
\bibitem [{\citenamefont {Ni}\ \emph {et~al.}(2015)\citenamefont {Ni},
  \citenamefont {Ding}, \citenamefont {Asta},\ and\ \citenamefont
  {Jin}}]{Ni2015}%
  \BibitemOpen
  \bibfield  {author} {\bibinfo {author} {\bibfnamefont {C.}~\bibnamefont
  {Ni}}, \bibinfo {author} {\bibfnamefont {H.}~\bibnamefont {Ding}}, \bibinfo
  {author} {\bibfnamefont {M.}~\bibnamefont {Asta}}, \ and\ \bibinfo {author}
  {\bibfnamefont {X.}~\bibnamefont {Jin}},\ }\href
  {http://dx.doi.org/10.1016/j.scriptamat.2015.07.028} {\bibfield  {journal}
  {\bibinfo  {journal} {Scripta Mater.}\ }\textbf {\bibinfo {volume} {109}},\
  \bibinfo {pages} {94 } (\bibinfo {year} {2015})}\BibitemShut {NoStop}%
\bibitem [{\citenamefont {Kasukabe}\ \emph {et~al.}(1993)\citenamefont
  {Kasukabe}, \citenamefont {Yamada}, \citenamefont {Lin},\ and\ \citenamefont
  {Bursill}}]{Kasukabe1993}%
  \BibitemOpen
  \bibfield  {author} {\bibinfo {author} {\bibfnamefont {Y.}~\bibnamefont
  {Kasukabe}}, \bibinfo {author} {\bibfnamefont {Y.}~\bibnamefont {Yamada}},
  \bibinfo {author} {\bibfnamefont {P.~J.}\ \bibnamefont {Lin}}, \ and\
  \bibinfo {author} {\bibfnamefont {L.~A.}\ \bibnamefont {Bursill}},\ }\href
  {\doibase http://dx.doi.org/10.1080/01418619308213983} {\bibfield  {journal}
  {\bibinfo  {journal} {Philos. Mag. A}\ }\textbf {\bibinfo {volume} {68}},\
  \bibinfo {pages} {587 } (\bibinfo {year} {1993})}\BibitemShut {NoStop}%
\bibitem [{\citenamefont {Pond}\ \emph {et~al.}(1995)\citenamefont {Pond},
  \citenamefont {Bacon},\ and\ \citenamefont {Serra}}]{Pond1995}%
  \BibitemOpen
  \bibfield  {author} {\bibinfo {author} {\bibfnamefont {R.~C.}\ \bibnamefont
  {Pond}}, \bibinfo {author} {\bibfnamefont {D.~J.}\ \bibnamefont {Bacon}}, \
  and\ \bibinfo {author} {\bibfnamefont {A.}~\bibnamefont {Serra}},\ }\href
  {\doibase 10.1080/09500839508240521} {\bibfield  {journal} {\bibinfo
  {journal} {Philos. Mag. Lett.}\ }\textbf {\bibinfo {volume} {71}},\ \bibinfo
  {pages} {275} (\bibinfo {year} {1995})}\BibitemShut {NoStop}%
\bibitem [{\citenamefont {Braisaz}\ \emph {et~al.}(1996)\citenamefont
  {Braisaz}, \citenamefont {Ruterana}, \citenamefont {Nouet}, \citenamefont
  {Serra}, \citenamefont {Komninou}, \citenamefont {Kehagias},\ and\
  \citenamefont {Karakostas}}]{Braisaz1996}%
  \BibitemOpen
  \bibfield  {author} {\bibinfo {author} {\bibfnamefont {T.}~\bibnamefont
  {Braisaz}}, \bibinfo {author} {\bibfnamefont {P.}~\bibnamefont {Ruterana}},
  \bibinfo {author} {\bibfnamefont {G.}~\bibnamefont {Nouet}}, \bibinfo
  {author} {\bibfnamefont {A.}~\bibnamefont {Serra}}, \bibinfo {author}
  {\bibfnamefont {P.}~\bibnamefont {Komninou}}, \bibinfo {author}
  {\bibfnamefont {T.}~\bibnamefont {Kehagias}}, \ and\ \bibinfo {author}
  {\bibfnamefont {T.}~\bibnamefont {Karakostas}},\ }\href {\doibase
  http://dx.doi.org/10.1080/095008396180056} {\bibfield  {journal} {\bibinfo
  {journal} {Philos. Mag. Lett.}\ }\textbf {\bibinfo {volume} {74}},\ \bibinfo
  {pages} {331 } (\bibinfo {year} {1996})}\BibitemShut {NoStop}%
\bibitem [{\citenamefont {Lay}\ and\ \citenamefont {Nouet}(1994)}]{Lay1994}%
  \BibitemOpen
  \bibfield  {author} {\bibinfo {author} {\bibfnamefont {S.}~\bibnamefont
  {Lay}}\ and\ \bibinfo {author} {\bibfnamefont {G.}~\bibnamefont {Nouet}},\
  }\href {\doibase http://dx.doi.org/10.1080/01418619408242947} {\bibfield
  {journal} {\bibinfo  {journal} {Philos. Mag. A}\ }\textbf {\bibinfo {volume}
  {70}},\ \bibinfo {pages} {261 } (\bibinfo {year} {1994})}\BibitemShut
  {NoStop}%
\bibitem [{\citenamefont {Braisaz}\ \emph {et~al.}(1997)\citenamefont
  {Braisaz}, \citenamefont {Ruterana},\ and\ \citenamefont
  {Nouet}}]{Braisaz1997}%
  \BibitemOpen
  \bibfield  {author} {\bibinfo {author} {\bibfnamefont {T.}~\bibnamefont
  {Braisaz}}, \bibinfo {author} {\bibfnamefont {P.}~\bibnamefont {Ruterana}}, \
  and\ \bibinfo {author} {\bibfnamefont {G.}~\bibnamefont {Nouet}},\ }\href
  {\doibase 10.1080/01418619708209962} {\bibfield  {journal} {\bibinfo
  {journal} {Phil. Mag. A}\ }\textbf {\bibinfo {volume} {76}},\ \bibinfo
  {pages} {63 } (\bibinfo {year} {1997})}\BibitemShut {NoStop}%
\bibitem [{\citenamefont {Zhang}\ \emph {et~al.}(2012)\citenamefont {Zhang},
  \citenamefont {Li}, \citenamefont {Wu}, \citenamefont {Zhu}, \citenamefont
  {Ma}, \citenamefont {Liu}, \citenamefont {Wang},\ and\ \citenamefont
  {Horstemeyer}}]{Zhang2012}%
  \BibitemOpen
  \bibfield  {author} {\bibinfo {author} {\bibfnamefont {X.~Y.}\ \bibnamefont
  {Zhang}}, \bibinfo {author} {\bibfnamefont {B.}~\bibnamefont {Li}}, \bibinfo
  {author} {\bibfnamefont {X.~L.}\ \bibnamefont {Wu}}, \bibinfo {author}
  {\bibfnamefont {Y.~T.}\ \bibnamefont {Zhu}}, \bibinfo {author} {\bibfnamefont
  {Q.}~\bibnamefont {Ma}}, \bibinfo {author} {\bibfnamefont {Q.}~\bibnamefont
  {Liu}}, \bibinfo {author} {\bibfnamefont {P.~T.}\ \bibnamefont {Wang}}, \
  and\ \bibinfo {author} {\bibfnamefont {M.~F.}\ \bibnamefont {Horstemeyer}},\
  }\href {\doibase 10.1016/j.scriptamat.2012.08.012} {\bibfield  {journal}
  {\bibinfo  {journal} {Scripta Mater.}\ }\textbf {\bibinfo {volume} {67}},\
  \bibinfo {pages} {862} (\bibinfo {year} {2012})}\BibitemShut {NoStop}%
\bibitem [{\citenamefont {Sun}\ \emph {et~al.}(2014)\citenamefont {Sun},
  \citenamefont {Zhang}, \citenamefont {Ren}, \citenamefont {Tu},\ and\
  \citenamefont {Liu}}]{Sun2014}%
  \BibitemOpen
  \bibfield  {author} {\bibinfo {author} {\bibfnamefont {Q.}~\bibnamefont
  {Sun}}, \bibinfo {author} {\bibfnamefont {X.~Y.}\ \bibnamefont {Zhang}},
  \bibinfo {author} {\bibfnamefont {Y.}~\bibnamefont {Ren}}, \bibinfo {author}
  {\bibfnamefont {J.}~\bibnamefont {Tu}}, \ and\ \bibinfo {author}
  {\bibfnamefont {Q.}~\bibnamefont {Liu}},\ }\href {\doibase
  http://dx.doi.org/10.1016/j.scriptamat.2014.07.012} {\bibfield  {journal}
  {\bibinfo  {journal} {Scripta Mater.}\ }\textbf {\bibinfo {volume} {90 -
  91}},\ \bibinfo {pages} {41 } (\bibinfo {year} {2014})}\BibitemShut {NoStop}%
\bibitem [{\citenamefont {Bacon}\ and\ \citenamefont
  {Serra}(1991)}]{Bacon1991}%
  \BibitemOpen
  \bibfield  {author} {\bibinfo {author} {\bibfnamefont {D.~J.}\ \bibnamefont
  {Bacon}}\ and\ \bibinfo {author} {\bibfnamefont {A.}~\bibnamefont {Serra}},\
  }\href {\doibase 10.1557/proc-238-73} {\bibfield  {journal} {\bibinfo
  {journal} {{MRS} Proc.}\ }\textbf {\bibinfo {volume} {238}},\ \bibinfo
  {pages} {73} (\bibinfo {year} {1991})}\BibitemShut {NoStop}%
\bibitem [{\citenamefont {Hirth}\ and\ \citenamefont {Pond}(1996)}]{Hirth1996}%
  \BibitemOpen
  \bibfield  {author} {\bibinfo {author} {\bibfnamefont {J.~P.}\ \bibnamefont
  {Hirth}}\ and\ \bibinfo {author} {\bibfnamefont {R.~C.}\ \bibnamefont
  {Pond}},\ }\href {\doibase http://dx.doi.org/10.1016/S1359-6454(96)00132-2}
  {\bibfield  {journal} {\bibinfo  {journal} {Acta Mater.}\ }\textbf {\bibinfo
  {volume} {44}},\ \bibinfo {pages} {4749 } (\bibinfo {year}
  {1996})}\BibitemShut {NoStop}%
\bibitem [{\citenamefont {Wang}\ \emph {et~al.}(2011)\citenamefont {Wang},
  \citenamefont {Beyerlein}, \citenamefont {Hirth},\ and\ \citenamefont
  {Tom{\'e}}}]{Wang2011}%
  \BibitemOpen
  \bibfield  {author} {\bibinfo {author} {\bibfnamefont {J.}~\bibnamefont
  {Wang}}, \bibinfo {author} {\bibfnamefont {I.~J.}\ \bibnamefont {Beyerlein}},
  \bibinfo {author} {\bibfnamefont {J.~P.}\ \bibnamefont {Hirth}}, \ and\
  \bibinfo {author} {\bibfnamefont {C.~N.}\ \bibnamefont {Tom{\'e}}},\ }\href
  {\doibase 10.1016/j.actamat.2011.03.024} {\bibfield  {journal} {\bibinfo
  {journal} {Acta Mater.}\ }\textbf {\bibinfo {volume} {59}},\ \bibinfo {pages}
  {3990} (\bibinfo {year} {2011})}\BibitemShut {NoStop}%
\bibitem [{\citenamefont {Li}\ \emph {et~al.}(2012)\citenamefont {Li},
  \citenamefont {El~Kadiri},\ and\ \citenamefont {Horstemeyer}}]{Li2012}%
  \BibitemOpen
  \bibfield  {author} {\bibinfo {author} {\bibfnamefont {B.}~\bibnamefont
  {Li}}, \bibinfo {author} {\bibfnamefont {H.}~\bibnamefont {El~Kadiri}}, \
  and\ \bibinfo {author} {\bibfnamefont {M.~F.}\ \bibnamefont {Horstemeyer}},\
  }\href {\doibase 10.1080/14786435.2011.637985} {\bibfield  {journal}
  {\bibinfo  {journal} {Philos. Mag.}\ }\textbf {\bibinfo {volume} {92}},\
  \bibinfo {pages} {1006} (\bibinfo {year} {2012})}\BibitemShut {NoStop}%
\bibitem [{\citenamefont {Khater}\ \emph {et~al.}(2013)\citenamefont {Khater},
  \citenamefont {Serra},\ and\ \citenamefont {Pond}}]{Khater2013}%
  \BibitemOpen
  \bibfield  {author} {\bibinfo {author} {\bibfnamefont {H.~A.}\ \bibnamefont
  {Khater}}, \bibinfo {author} {\bibfnamefont {A.}~\bibnamefont {Serra}}, \
  and\ \bibinfo {author} {\bibfnamefont {R.~C.}\ \bibnamefont {Pond}},\ }\href
  {\doibase 10.1080/14786435.2013.769071} {\bibfield  {journal} {\bibinfo
  {journal} {Philos. Mag.}\ }\textbf {\bibinfo {volume} {93}},\ \bibinfo
  {pages} {1279} (\bibinfo {year} {2013})}\BibitemShut {NoStop}%
\bibitem [{\citenamefont {Hartley}\ and\ \citenamefont
  {Mishin}(2005)}]{Hartley2005}%
  \BibitemOpen
  \bibfield  {author} {\bibinfo {author} {\bibfnamefont {C.~S.}\ \bibnamefont
  {Hartley}}\ and\ \bibinfo {author} {\bibfnamefont {Y.}~\bibnamefont
  {Mishin}},\ }\href {\doibase http://dx.doi.org/10.1016/j.actamat.2004.11.027}
  {\bibfield  {journal} {\bibinfo  {journal} {Acta Mater.}\ }\textbf {\bibinfo
  {volume} {53}},\ \bibinfo {pages} {1313 } (\bibinfo {year}
  {2005})}\BibitemShut {NoStop}%
\bibitem [{\citenamefont {Cai}\ \emph {et~al.}(2003)\citenamefont {Cai},
  \citenamefont {Bulatov}, \citenamefont {Chang}, \citenamefont {Li},\ and\
  \citenamefont {Yip}}]{Cai2003}%
  \BibitemOpen
  \bibfield  {author} {\bibinfo {author} {\bibfnamefont {W.}~\bibnamefont
  {Cai}}, \bibinfo {author} {\bibfnamefont {V.~V.}\ \bibnamefont {Bulatov}},
  \bibinfo {author} {\bibfnamefont {J.}~\bibnamefont {Chang}}, \bibinfo
  {author} {\bibfnamefont {J.}~\bibnamefont {Li}}, \ and\ \bibinfo {author}
  {\bibfnamefont {S.}~\bibnamefont {Yip}},\ }\href {\doibase
  10.1080/0141861021000051109} {\bibfield  {journal} {\bibinfo  {journal}
  {Philos. Mag.}\ }\textbf {\bibinfo {volume} {83}},\ \bibinfo {pages} {539}
  (\bibinfo {year} {2003})}\BibitemShut {NoStop}%
\bibitem [{\citenamefont {Rodney}\ \emph {et~al.}(2017)\citenamefont {Rodney},
  \citenamefont {Ventelon}, \citenamefont {Clouet}, \citenamefont
  {Pizzagalli},\ and\ \citenamefont {Willaime}}]{Rodney2017}%
  \BibitemOpen
  \bibfield  {author} {\bibinfo {author} {\bibfnamefont {D.}~\bibnamefont
  {Rodney}}, \bibinfo {author} {\bibfnamefont {L.}~\bibnamefont {Ventelon}},
  \bibinfo {author} {\bibfnamefont {E.}~\bibnamefont {Clouet}}, \bibinfo
  {author} {\bibfnamefont {L.}~\bibnamefont {Pizzagalli}}, \ and\ \bibinfo
  {author} {\bibfnamefont {F.}~\bibnamefont {Willaime}},\ }\href {\doibase
  10.1016/j.actamat.2016.09.049} {\bibfield  {journal} {\bibinfo  {journal}
  {Acta Mater.}\ ,\ \bibinfo {pages} {633 }} (\bibinfo {year}
  {2017})}\BibitemShut {NoStop}%
\bibitem [{\citenamefont {Mura}(1987)}]{Mura1987}%
  \BibitemOpen
  \bibfield  {author} {\bibinfo {author} {\bibfnamefont {T.}~\bibnamefont
  {Mura}},\ }\href@noop {} {\emph {\bibinfo {title} {Micromechanics of Defects
  in Solids}}}\ (\bibinfo  {publisher} {Kluwer Academic},\ \bibinfo {year}
  {1987})\BibitemShut {NoStop}%
\bibitem [{\citenamefont {Moulinec}\ and\ \citenamefont
  {Suquet}(1998)}]{Moulinec1998}%
  \BibitemOpen
  \bibfield  {author} {\bibinfo {author} {\bibfnamefont {H.}~\bibnamefont
  {Moulinec}}\ and\ \bibinfo {author} {\bibfnamefont {P.}~\bibnamefont
  {Suquet}},\ }\href {\doibase http://dx.doi.org/10.1016/S0045-7825(97)00218-1}
  {\bibfield  {journal} {\bibinfo  {journal} {Comput. Methods Appl. Mech.
  Engrg.}\ }\textbf {\bibinfo {volume} {157}},\ \bibinfo {pages} {69 }
  (\bibinfo {year} {1998})}\BibitemShut {NoStop}%
\bibitem [{\citenamefont {Dupeux}\ and\ \citenamefont
  {Bonnet}(1980)}]{Dupeux1980}%
  \BibitemOpen
  \bibfield  {author} {\bibinfo {author} {\bibfnamefont {M.}~\bibnamefont
  {Dupeux}}\ and\ \bibinfo {author} {\bibfnamefont {R.}~\bibnamefont
  {Bonnet}},\ }\href {\doibase 10.1016/0001-6160(80)90150-9} {\bibfield
  {journal} {\bibinfo  {journal} {Acta Metall.}\ }\textbf {\bibinfo {volume}
  {28}},\ \bibinfo {pages} {721} (\bibinfo {year} {1980})}\BibitemShut
  {NoStop}%
\bibitem [{\citenamefont {Rapperport}(1959)}]{Rapperport1959}%
  \BibitemOpen
  \bibfield  {author} {\bibinfo {author} {\bibfnamefont {K.~E.~J.}\
  \bibnamefont {Rapperport}},\ }\href {\doibase 10.1016/0001-6160(59)90018-5}
  {\bibfield  {journal} {\bibinfo  {journal} {Acta Metall.}\ }\textbf {\bibinfo
  {volume} {7}},\ \bibinfo {pages} {254} (\bibinfo {year} {1959})}\BibitemShut
  {NoStop}%
\bibitem [{\citenamefont {Rapperport}\ and\ \citenamefont
  {Hartley}(1960)}]{Rapperport1960}%
  \BibitemOpen
  \bibfield  {author} {\bibinfo {author} {\bibfnamefont {K.~E.~J.}\
  \bibnamefont {Rapperport}}\ and\ \bibinfo {author} {\bibfnamefont {C.~S.}\
  \bibnamefont {Hartley}},\ }\href@noop {} {\bibfield  {journal} {\bibinfo
  {journal} {Trans. AIME}\ }\textbf {\bibinfo {volume} {218}},\ \bibinfo
  {pages} {869} (\bibinfo {year} {1960})}\BibitemShut {NoStop}%
\bibitem [{\citenamefont {Paton}\ and\ \citenamefont
  {Backofen}(1970)}]{Paton1970}%
  \BibitemOpen
  \bibfield  {author} {\bibinfo {author} {\bibfnamefont {N.~E.}\ \bibnamefont
  {Paton}}\ and\ \bibinfo {author} {\bibfnamefont {W.~A.}\ \bibnamefont
  {Backofen}},\ }\href {\doibase 10.1007/BF03037822} {\bibfield  {journal}
  {\bibinfo  {journal} {Metall. Trans.}\ }\textbf {\bibinfo {volume} {1}},\
  \bibinfo {pages} {2839 } (\bibinfo {year} {1970})}\BibitemShut {NoStop}%
\bibitem [{\citenamefont {Akhtar}(1973)}]{Akhtar1973a}%
  \BibitemOpen
  \bibfield  {author} {\bibinfo {author} {\bibfnamefont {A.}~\bibnamefont
  {Akhtar}},\ }\href {\doibase 10.1016/0022-3115(73)90189-X} {\bibfield
  {journal} {\bibinfo  {journal} {J. Nucl. Mater.}\ }\textbf {\bibinfo {volume}
  {47}},\ \bibinfo {pages} {79} (\bibinfo {year} {1973})}\BibitemShut {NoStop}%
\bibitem [{\citenamefont {Jensen}\ and\ \citenamefont
  {Backofen}(1972)}]{Jensen1972}%
  \BibitemOpen
  \bibfield  {author} {\bibinfo {author} {\bibfnamefont {J.~A.}\ \bibnamefont
  {Jensen}}\ and\ \bibinfo {author} {\bibfnamefont {W.~A.}\ \bibnamefont
  {Backofen}},\ }\href {\doibase 10.1179/cmq.1972.11.1.39} {\bibfield
  {journal} {\bibinfo  {journal} {Can. Metall. Q.}\ }\textbf {\bibinfo {volume}
  {11}},\ \bibinfo {pages} {39} (\bibinfo {year} {1972})}\BibitemShut {NoStop}%
\bibitem [{\citenamefont {Henkelman}\ and\ \citenamefont
  {Jonsson}(2000)}]{Henkelman2000}%
  \BibitemOpen
  \bibfield  {author} {\bibinfo {author} {\bibfnamefont {G.}~\bibnamefont
  {Henkelman}}\ and\ \bibinfo {author} {\bibfnamefont {H.}~\bibnamefont
  {Jonsson}},\ }\href {\doibase 10.1063/1.1323224} {\bibfield  {journal}
  {\bibinfo  {journal} {J. Chem. Phys.}\ }\textbf {\bibinfo {volume} {113}},\
  \bibinfo {pages} {9978 } (\bibinfo {year} {2000})}\BibitemShut {NoStop}%
\bibitem [{\citenamefont {Combe}\ \emph {et~al.}(2016)\citenamefont {Combe},
  \citenamefont {Mompiou},\ and\ \citenamefont {Legros}}]{Combe2016}%
  \BibitemOpen
  \bibfield  {author} {\bibinfo {author} {\bibfnamefont {N.}~\bibnamefont
  {Combe}}, \bibinfo {author} {\bibfnamefont {F.}~\bibnamefont {Mompiou}}, \
  and\ \bibinfo {author} {\bibfnamefont {M.}~\bibnamefont {Legros}},\ }\href
  {\doibase 10.1103/PhysRevB.93.024109} {\bibfield  {journal} {\bibinfo
  {journal} {Phys. Rev. B}\ }\textbf {\bibinfo {volume} {93}},\ \bibinfo
  {pages} {024109} (\bibinfo {year} {2016})}\BibitemShut {NoStop}%
\bibitem [{\citenamefont {Proville}\ \emph {et~al.}(2013)\citenamefont
  {Proville}, \citenamefont {Ventelon},\ and\ \citenamefont
  {Rodney}}]{Proville2013}%
  \BibitemOpen
  \bibfield  {author} {\bibinfo {author} {\bibfnamefont {L.}~\bibnamefont
  {Proville}}, \bibinfo {author} {\bibfnamefont {L.}~\bibnamefont {Ventelon}},
  \ and\ \bibinfo {author} {\bibfnamefont {D.}~\bibnamefont {Rodney}},\ }\href
  {\doibase 10.1103/PhysRevB.87.144106} {\bibfield  {journal} {\bibinfo
  {journal} {Phys. Rev. B}\ }\textbf {\bibinfo {volume} {87}},\ \bibinfo
  {pages} {144106} (\bibinfo {year} {2013})}\BibitemShut {NoStop}%
\bibitem [{\citenamefont {Kraych}\ \emph {et~al.}(2016)\citenamefont {Kraych},
  \citenamefont {Carrez}, \citenamefont {Hirel}, \citenamefont {Clouet},\ and\
  \citenamefont {Cordier}}]{Kraych2016}%
  \BibitemOpen
  \bibfield  {author} {\bibinfo {author} {\bibfnamefont {A.}~\bibnamefont
  {Kraych}}, \bibinfo {author} {\bibfnamefont {P.}~\bibnamefont {Carrez}},
  \bibinfo {author} {\bibfnamefont {P.}~\bibnamefont {Hirel}}, \bibinfo
  {author} {\bibfnamefont {E.}~\bibnamefont {Clouet}}, \ and\ \bibinfo {author}
  {\bibfnamefont {P.}~\bibnamefont {Cordier}},\ }\href {\doibase
  10.1103/PhysRevB.93.014103} {\bibfield  {journal} {\bibinfo  {journal} {Phys.
  Rev. B}\ }\textbf {\bibinfo {volume} {93}},\ \bibinfo {pages} {014103}
  (\bibinfo {year} {2016})}\BibitemShut {NoStop}%
\end{thebibliography}%
\end{document}